\documentclass[aps,pra,reprint,groupaddress]{revtex4-1}
\usepackage[utf8]{inputenc}
\usepackage{bm}
\usepackage{enumerate}
\usepackage{graphicx}
\usepackage{amsmath}
\usepackage{amssymb}
\usepackage{cancel}
\usepackage{xcolor}
\usepackage{verbatim}
\usepackage{hyperref}
\usepackage{lipsum}
\usepackage{bbold}
\usepackage{comment}
\usepackage{scalerel}
\hypersetup{colorlinks=true,
linkcolor=blue,
citecolor=blue,
urlcolor=blue,
filecolor=blue
}

\def\bra#1{\langle{#1}|}
\def\ket#1{|{#1}\rangle}

\newcommand{\bs}{\boldsymbol}
\newcommand{\ua}{\uparrow}
\newcommand{\da}{\downarrow}

\begin{abstract}
Spin--orbit interaction (SOI) plays a fundamental role in many low-dimensional semiconductor and hybrid quantum devices.
In the rapidly evolving field of semiconductor spin qubits, SOI is an essential ingredient that can allow for ultrafast qubit control.
The exact manifestation of SOI in a given device is, however, often both hard to predict theoretically and probe experimentally.
Here, we develop a detailed theoretical connection between the leakage current through a double quantum dot in Pauli spin blockade and the underlying SOI in the system.
We present a general analytic expression for the leakage current, which allows to connect experimentally observable features to both the magnitude and orientation of the effective spin--orbit field acting on the moving carriers.
Motivated by the large recent interest in hole-based quantum devices, we further zoom in on the case of Pauli blockade of hole spins, assuming a strong transverse confinement potential.
In this limit we also find an analytic expression for the current at low external magnetic field, that includes the effect of hyperfine coupling of the hole spins to randomly fluctuating nuclear spin baths.
This result can be used to extract detailed information about both hyperfine and spin--orbit coupling parameters for hole spins in devices with a significant fraction of non-zero nuclear spins.
\end{abstract}

\begin{document}

\title{Probing details of spin--orbit coupling through Pauli spin blockade}

\author{Jørgen Holme Qvist}
\affiliation{Center for Quantum Spintronics, Department of Physics, Norwegian University of Science and Technology, NO-7491 Trondheim, Norway}

\author{Jeroen Danon}
\affiliation{Center for Quantum Spintronics, Department of Physics, Norwegian University of Science and Technology, NO-7491 Trondheim, Norway}

\date{\today}

\maketitle

\section{Introduction}

Spin--orbit interaction (SOI) couples the spin degree of freedom of a charge carrier moving in an electromagnetic field to its momentum.
This interaction is an essential ingredient for many semiconductor-based quantum devices.
In semiconductor-superconductor hybrid structures, SOI plays a crucial role for the realization of Majorana bound states \cite{sau2010generic,lutchyn2010majorana,oreg2010helical,leijnse2012introduction,beenakker2013search,Lutchyn2018May}, with potential applications in topologically protected quantum computation~\cite{kitaev2003fault,nayak2008non,sarma2015majorana,stanescu2016introduction}.
For spin-based quantum technologies SOI enables spin manipulation via electric control, 
allowing for enhanced spin-cavity couplings \cite{Petersson2012Oct,Samkharadze2018,burkard2020superconductor} and electric dipole spin resonance \cite{Rashba2003Sep,Zutic2004Apr,Nadj-Perge2010Dec}.

In the field of semiconductor spin qubits~\cite{Loss1998,Hanson2007Oct,Kloeffel2013Mar,Chatterjee2021Mar,burkard2021semiconductor}, the electric control over spin provided by SOI enables fast qubit operation \cite{Golovach2006Oct,Flindt2006Dec}, while also being a source of qubit decoherence and relaxation \cite{Khaetskii2000,Khaetskii2001,golovach2004phonon}. 
Lately, there has been substantial progress with Si- and Ge-based spin qubits that use the spin of valence-band holes instead of conduction-band electrons~\cite{Maurand2016,Watzinger2018,Vukusic2018,Crippa2019,Scappucci2020,Jirovec2021,Lawrie2021,Hendrickx2021}.
The $p$-type nature of the valence band leads to a mixing of the orbital and spin degrees of freedom of the carriers, yielding a potentially strong effective SOI that depends on the details of the confinement.
This can give rise to several interesting phenomena such as a highly anisotropic and electrically tunable $g$-tensor \cite{watzinger2016,Brauns2016,Bogan2017,Lu2017,Crippa2018,Marcellina2018,Gradl2018,hofmann2019ArXiV,Miller2021,Froning2021,Liles2021Dec,Qvist2022}, and it could also allow for very fast spin-qubit manipulation~\cite{ares2013,Kloeffel2011,Kloeffel2018,Bosco2021,Bosco2021a,Froning2021a,Wang2021}. 

Despite SOI being crucial for the working of many semiconductor quantum devices, its exact manifestation for a given system is often hard to predict or deduce from experiments.
This is partly a result of the total SOI having  often several, qualitatively different contributions in strongly confined systems~\cite{winkler2003,Marcellina2017}. Common contributions are Rashba terms stemming from structural inversion asymmetry, e.g., created by a confining potential, and Dresselhaus terms originating from the lack of a crystallographic inversion center in semiconductors with zinc-blende structure.
In addition to this, both the so-called dipolar SOI~\cite{Philippopoulos2020Aug} and strain-induced mixing of different hole states~\cite{Terrazos2021} can strongly affect the total effective SOI for valence-band spins.

In an experiment, the relevant spin--orbit parameters often emerge on a phenomenological level as an effective spin--orbit field that acts on the moving carriers.
The manifestation of this field can be probed using several different approaches, the most common ones being dispersive gate sensing \cite{Han2022Mar} and current measurements as a function of the orientation of an externally applied magnetic field~\cite{Wang2016Dec,Wang2018Aug,Marx2020Mar,Sala2021}. 
Since such measurements are the most straightforward way to access the details of the effective spin--orbit field, it is essential to develop a thorough understanding of the connection between the experimentally accessible quantities and the underlying SOI.

In this paper, we focus on a double quantum dot tuned to the regime of Pauli spin blockade, where the most important effect of SOI is that it effectively allows for interdot tunneling accompanied by a spin rotation~\cite{danon2009}, which fundamentally changes the nature of the blockade.
We theoretically investigate the leakage current through the system, focusing on its dependence on the details of the SOI.
Building on the approach of Refs.~\cite{jouravlev2006,danon2013}, which found expressions for the current in absence of SOI, we derive a general analytic expression for the leakage current including SOI.
Based on this result we present a straightforward connection between the details of the emerging spin--orbit field and the dependence of the current on experimentally tunable parameters.
We also discuss the role of the hyperfine interaction between the localized spins and randomly fluctuating nuclear spins of the host material, most relevant for devices based on III-V semiconductors.
For the case of hole-based systems defined in a strongly confined two-dimensional hole gas we present an analytic expression that captures the combined effects on the leakage current of SOI and coupling to randomly polarized nuclear spin baths.
Comparing this expression to an experimentally measured current could reveal details about both the effective nuclear fields on the dots and the spin--orbit field in the system.

The rest of the paper is organized as follows. In Sec.~\ref{sec:model} we present our model Hamiltonian used to describe the double-dot system.
In Sec.~\ref{sec:current} we then derive an analytic expression for the leakage current as a function of arbitrarily oriented spin--orbit and Zeeman fields.
Based on this expression we characterize special points in parameter space where the current vanishes.
In Sec.~\ref{sec:soi} we consider the collection of these stopping points and we present straightforward connections between clear features of the current (such as sharp minima) and the orientation and magnitude of the effective spin--orbit field in the system. In Sec.~\ref{sec:independently} we assume fully controllable Zeeman fields on the two dots and in Sec.~\ref{sec:homogeneousfield} we assume a homogeneously applied field but allow for additional random nuclear fields on the dots, focusing on hole-spin systems with strong transverse confinement.

\section{Model}
\label{sec:model}

The system we consider consists of two tunnel coupled quantum dots that both are connected to a lead, as illustrated in Fig.~\ref{fig:figure}. 
We assume that the system is tuned close to the (1,1)--(0,2) charge transition, where $(n,m)$ indicates a state with $n(m)$ excess charges on the left(right) dot, which can be either electrons or holes.
Applying a voltage bias between the two leads can then induce a current to run through the double dot, say from the left to the right lead. 
Assuming a large on-site orbital level splitting (typically $\sim$~meV) compared to the applied bias voltage, states involving excited orbitals can be disregarded, and the Pauli exclusion principle then dictates that the two charges in the (0,2) configuration must be in a spin-singlet state, $\ket{S_{02}}$.
In the (1,1) charge configuration all four spin states are accessible, three triplets $\ket{T_{\pm,0}}$ and one singlet $\ket{S}$.
This can lead to a so-called spin blockade, where the system is stuck in one of the (1,1) triplet states, which cannot transition to $\ket{S_{02}}$.

We include two spin-mixing ingredients that can modify or lift this blockade.
Firstly, each of the two dots experiences a Zeeman field, $\bs B_{L,R}$, which we allow to be different on the two dots.
These effective magnetic fields can originate from an externally applied field, nearby on-chip micromagnets, or hyperfine interaction between the localized spins and the nuclear spins of the host material.
Secondly, we also allow for strong spin--orbit coupling.
This can result in spin flips during tunneling between the dots, but it can also renormalize the $g$-tensors on the two dots, potentially contributing to a difference in the effective Zeeman fields on the two dots.

Focusing on the five levels mentioned above, we describe the system with a simple model Hamiltonian,
\begin{equation}
    H = H_e + H_t + H_B.\label{eq:hamtot}
\end{equation}
Here
\begin{equation}
    {H}_e = -\delta \ket{S_{02}}\bra{S_{02}}
\end{equation}
accounts for the relative detuning $\delta$ of the four (1,1) states with respect to the (0,2) singlet.
The interdot tunnel coupling is described by
\begin{align}
    {H}_t =
    t_s\ket{S}\bra{S_{02}} 
    + i\bs t_\text{so}\cdot\ket{\bs T}\bra{S_{02}}
    + \text{H.c.},    \label{eq:ham_tun}
\end{align}
where $\ket{\bs T} = \{ \ket{T_x}, \ket{T_y}, \ket{T_z} \}$ is the vector of unpolarized triplet states along the three orthogonal coordinate axes~\cite{danon2009}.
The first term in $H_t$ accounts for spin-conserving tunneling, whereas the second term parametrizes the effect of spin--orbit interaction on the interdot tunneling, effectively yielding spin-non-conserving tunneling terms.
The magnitude and orientation of the vector $\bs t_\text{so}$ depend on microscopic details of the spin--orbit interaction.
Finally, due to the singlet nature of $\ket{S_{02}}$, magnetic fields only yield a Zeeman effect within the (1,1) subspace, which we describe by
\begin{equation}
    H_B = \frac{1}{2}\left[\left( \bs B_{L}\cdot\bs\sigma_L \right)\otimes\mathbb{1}_R + \mathbb{1}_L\otimes\left( \bs B_{R}\cdot\bs\sigma_R \right)\right],
    \label{eq:zeeman}
\end{equation}
with $\bs \sigma_{L(R)}$ being the vector of Pauli matrices acting on the left(right) spin and $\bs B_{L(R)}$ being the total Zeeman field on the left(right) dot.
These fields can contain a contribution from externally applied magnetic fields as well as the Overhauser fields that are due to hyperfine interaction with the spinful nuclei in each dot.

\section{Leakage current}
\label{sec:current}

The current through the double dot, and thus the degree of spin blockade, is governed by an interplay between the structure of the coupling Hamiltonian (\ref{eq:ham_tun}) and the degree of spin mixing within the (1,1) subspace due to the fields $\bs B_{L,R}$.
Because of the resulting complexity it will be convenient to perform a basis transformation which makes the Hamiltonian take a simple form, from which the current can be calculated analytically.

\begin{figure}[t]
	\centering
	\includegraphics[width=\linewidth]{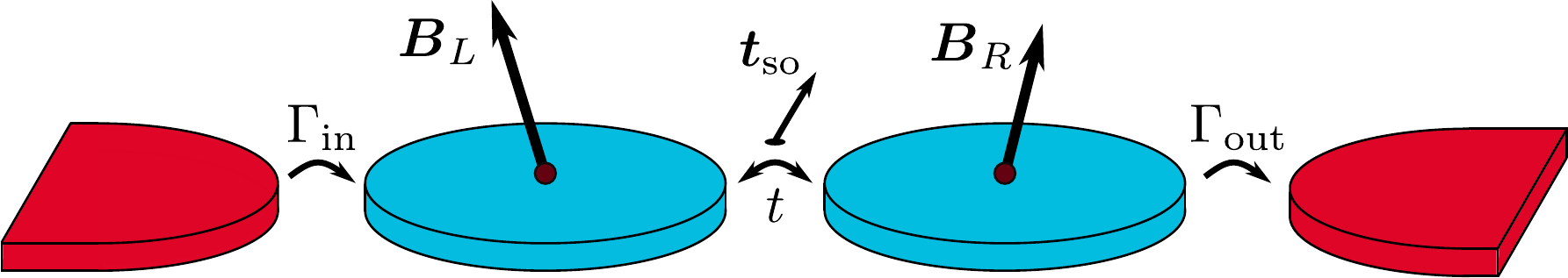} 
	\caption{Illustration of the two tunnel coupled quantum dots connected to two leads, showing the orientation of the different fields: The spin-orbit vector is assumed to be pointing along $\hat{z}$, whereas the Zeeman fields $E_{Z,i}$ and Overhauser fields $K_i$ are arbitrary. }
	\label{fig:figure}
\end{figure}

The first step is to define the $z$-direction of our coordinate system to point along $\bs t_{\rm so}$.
This rotates the coupling Hamiltonian into $H_t = t_s\ket{S}\bra{S_{02}} + i t_\text{so} \ket{T_0}\bra{S_{02}} + \text{H.c.}$, where $\ket{T_0} = \frac{1}{\sqrt 2}\big[ \ket{\ua\da} + \ket{\da\ua}\big]$ is the (usual) unpolarized spin triplet along $\hat z$ and $t_\text{so}$ is the magnitude of the spin--orbit vector $\bs t_{\rm so}$.
We then introduce a dimensionless parameter $\eta=\arctan\left[t_\text{so}/t_s\right]$ that parameterizes the relative strength of the spin--orbit-induced tunnel coupling and apply a basis transformation to all (1,1) states
\begin{equation}
    \ket{\tilde \psi} = e^{i\frac{\eta}{2}(\sigma_L^z - \sigma_R^z)}\ket{\psi}.
\end{equation}
In this new basis we find that $\ket{\tilde S} = \cos\eta\,\ket{S} + i\sin\eta\,\ket{T_0}$ is a ``bright'' state that is coupled to $\ket{S_{02}}$ with strength $t\equiv\sqrt{t_s^2+t_\text{so}^2}$, and $\ket{\tilde T_0} = i\sin\eta\,\ket{S} + \cos\eta\,\ket{T_0}$ is a ``dark'' state that is not coupled;
the polarized triplet states $\ket{\tilde T_\pm} = \ket{T_\pm}$ are unchanged by the transformation.
Therefore, in the new basis only one (1,1) state is coupled to $\ket{S_{02}}$, the price to pay being that the transformed Zeeman Hamiltonian $e^{-i\frac{\eta}{2}(\sigma_L^z - \sigma_R^z)}H_Be^{i\frac{\eta}{2}(\sigma_L^z - \sigma_R^z)}$ acquired an $\eta$-dependence and now incorporates all spin--orbit effects included in our model.
The transformation thus gauges away the spin--orbit interaction, yielding a Hamiltonian that can be mapped exactly to the case without spin--orbit coupling ($\bs t_{\rm so}=\bs 0$), simply by redefining the two effective Zeeman fields.
For the case without spin--orbit interaction steady-state expressions for the current have been derived before~\cite{jouravlev2006,danon2013} and one can thus apply a similar approach to include spin--orbit coupling.

We assume the system to be tuned to the open regime, where the couplings to the reservoirs, characterized by the tunneling rates $\Gamma_{\rm in,out}$ (see Fig.~\ref{fig:figure}), are the largest relevant energy scales.
This ensures that the sequential tunneling process $(0,2) \to (0,1) \to (1,1)$ is effectively instantaneous, and the interesting dynamics happen during the transition $(1,1) \to (0,2)$ which involves only the five levels we included in the Hamiltonian (\ref{eq:hamtot}).
An analytical expression for the current is then obtained by solving the steady-state Master equation (see App.~\ref{app:mastser_eq} for more details). In the limit $\Gamma\gg\delta,t,B_{L,R}$ we find the relatively compact expression (setting $\hbar=1$ from here on),
\begin{equation}
    \frac{I}{e\Gamma_s} = \frac{|e^{2i\eta} B_R^-B_L^z-B_L^-B_R^z|^2 + {\rm Im} \{ e^{2i\eta}B_R^-B_L^+\}^2 }{\Gamma_s^2Q_+^2\left[ 3 + \frac{16 Q_+^2Q_-^2}{(B_L^2-B_R^2)^2} \right] + B_L^2B_R^2},
    \label{eq:current}
\end{equation}
where we used $B_{L,R} = |\bs B_{L,R}|$ and we introduced the rate $\Gamma_s \equiv t^2/\Gamma$, which sets the scale of the effective decay rate of the (1,1) states.
We also introduced the notations $B^\pm = B^x \pm iB^y$ and
\begin{align}
    Q_\pm^2 = {} & {} {\rm Re}\{\tfrac{1}{2}e^{i\eta}(B_L^+ \pm B_R^-)\}^2 \nonumber\\
    {} & {} + {\rm Im}\{\tfrac{1}{2}e^{i\eta}(B_L^+ \mp  B_R^-)\}^2 +\tfrac{1}{4}(B_L^z \pm B_R^z)^2.
\end{align}
This expression thus describes the current through a double quantum dot in the spin-blockade regime, including the effect of spin--orbit coupling and two possibly different Zeeman fields on the two dots.

Eq.~(\ref{eq:current}) is the most important analytic result of this work; it generalizes the result presented in Ref.~\cite{danon2013}, by including arbitrarily oriented non-spin-conserving interdot tunneling processes.
The relative importance of these processes is described by the parameter $\eta = \arctan[ t_{\rm so}/t_s]$, so that $e^{i\eta} = (t_s/t) + i(t_{\rm so}/t)$, and the direction of the vector $\bs t_{\rm so}$ is encoded in the choice of coordinate system, by defining the $z$-direction along $\bs t_{\rm so}$.

\begin{figure}[t]
	\centering
	\includegraphics[width=\linewidth]{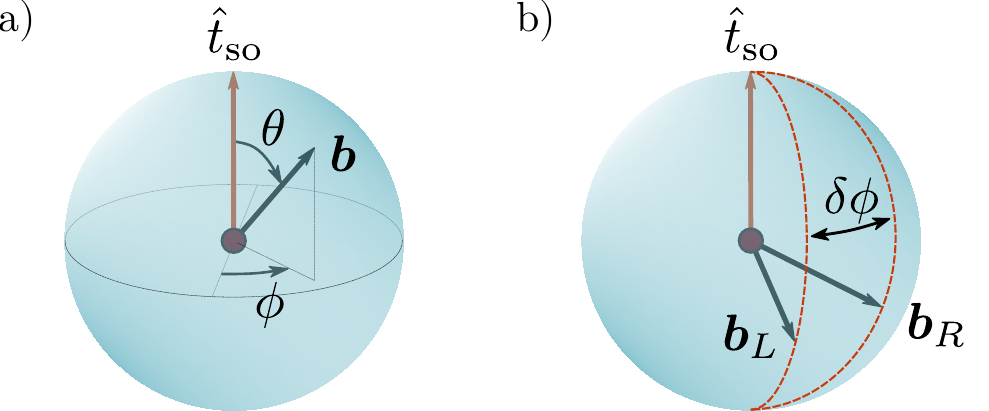} 
	\caption{
	(a) For two Zeeman fields with the same orientation $\bs b = \bs B/B$ but different magnitudes the current only vanishes when the two fields are oriented along the spin--orbit vector marked in red.
	(b) Tuning the Zeeman fields away from the spin--orbit vector, the current vanish along the red lines on the sphere where the relative orientation of the two Zeeman vectors ${\bs b}_{L,R}$ satisfy $\delta\phi = 2\eta$.
	}
	\label{fig:plot2}
\end{figure}

From Eq.~(\ref{eq:current}) we can identify special configurations of $\bs B_{L,R}$ for which the current vanishes, so-called ``stopping points''~\cite{jouravlev2006,danon2013}. 
We find four of such points:
(i) The first arises when the magnitude of the two Zeeman fields is equal, $B_{L}=B_R$, making the term $16Q_+^2Q_-^2/(B_L^2-B_R^2)^2$ in the denominator diverge.
The blockade at this point can be understood from considering the (1,1) states in the basis of spin up and down along the local fields on the left and right dot.
In this basis, the two states $\ket{\ua\da}$ and $\ket{\da\ua}$ are both eigenstates of $H_B$ with zero total Zeeman energy.
This means that they can be rearranged into a bright and a dark state (again in terms of coupling to $\ket{S_{02}}$) and the system will thus get blocked in the dark state.
(ii) The three other points are obtained for field configurations where the numerator in Eq.~(\ref{eq:current}) vanishes.
One configuration for which this happens is when either of the two fields is zero, $B_{L,R} = 0$, resulting in two doubly degenerate subspaces which can again be rearranged in dark and bright states.
(iii) The numerator also vanishes when both fields are parallel or antiparallel to the spin--orbit vector $\bs t_\text{so}$, i.e., $B_{L,R}^\pm = 0$.
In this case the two triplets $\ket{T_\pm}$ are eigenstates of $H_B$ that are not coupled to $\ket{S_{02}}$, resulting in a blockade of the current.
(iv) The last stopping point occurs when $e^{2i\eta} = B_L^-B_R^z/B_R^-B_L^z$.
Writing the two fields in spherical coordinates $\{B,\theta,\phi\}$, where $\theta=0$ corresponds to the $z$-direction (which is aligned with $\bs t_{\rm so}$), as illustrated in Fig.~\ref{fig:plot2}(a), we see that this condition corresponds to having $\phi_R-\phi_L=2\eta$ if $\theta_L = \theta_R$ and $\phi_R-\phi_L=2\eta + \pi$ if $\theta_L = \pi - \theta_R$.
This configuration corresponds to the two fields having the same ``latitude'' but a relative azimuthal angle of $\delta\phi = 2\eta$, as illustrated in Fig.~\ref{fig:plot2}(b) (or one of the two fields can have an overall minus sign compared to this situation).
This stopping point can be understood from considering the non-spin-conserving tunneling that underlies the coupling Hamiltonian (\ref{eq:ham_tun}):
$H_t$ can be interpreted as being a projection to our five-level basis of the general non-spin-conserving tunneling Hamiltonian
\begin{equation}
    H_t = \frac{1}{\sqrt 2}\hat c^\dagger_{L,\alpha} [ t_s \mathbb{1} + i\bs t_{\rm so}\cdot\bs \sigma ]_{\alpha\beta} \hat c_{R,\beta} + {\rm H.c.},
\end{equation}
where $\hat c^\dagger_{L(R),\sigma}$ is the creation operator of a charge with spin $\sigma$ on the left(right) dot.
With the $z$-axis oriented along $\bs t_{\rm so}$ we see that this tunneling Hamiltonian reduces to $H_t = \frac{1}{\sqrt 2}\hat c^\dagger_{L,\alpha} [ t\, e^{i\eta\sigma_z} ]_{\alpha\beta} \hat c_{R,\beta} + {\rm H.c.}$, which describes charge tunneling with amplitude $t$ that is accompanied by a $z$-rotation of the spin over an angle of $\pm 2\eta$ (depending on the direction of tunneling).
With this in mind we understand that the eigenstate of $H_B$ where both spins are aligned with (or exactly opposite to) two local fields that have a relative azimuthal angle of $2\eta$ will evolve during the interdot tunneling into a fully polarized spin-1 state, which has no overlap with $\ket{S_{02}}$.

\section{Effects of spin--orbit coupling}
\label{sec:soi}

\subsection{Independently controllable Zeeman fields}
\label{sec:independently}

The collection of stopping points provides a potentially useful tool for characterizing the spin--orbit interaction in a double-dot system, allowing to identify both the orientation and magnitude of the spin--orbit tunneling vector $\bs t_{\rm so}$.
Assuming that one has full control over the two Zeeman fields on the dots, either through local control of the applied magnetic fields or, e.g., via local manipulation of the $g$-tensor, one can in principle map out all the stopping points discussed above.

If one makes sure that the two Zeeman fields are both non-zero and have different magnitudes, then only the last two stopping points will be probed.
In this case, the orientation of the spin--orbit vector (up to a sign) can be identified from making the two Zeeman fields parallel to each other and finding the field orientation for which the current vanishes, i.e., by probing stopping point (iii)~\footnote{The current also vanishes for $\theta=\pi/2$ and $\eta=\pi/2$, where the couplings between $\ket{S_{02}}$ and $\{\ket{\ua\da},\ket{\da\ua}\}$ (in the eigenbasis of the local fields) are zero. However, the case $\eta=\pi/2$ corresponds to the extreme case where $t_{\rm so}=t$ and $t_s=0$, which we will ignore here.}.
Knowing the orientation of the spin--orbit vector, its magnitude can then be found by identifying stopping points of type (iv):
One tilts both fields away from $\bs t_{\rm so}$, in any direction, and then one rotates one of the fields along $\bs t_{\rm so}$ while measuring the leakage current.
From the point where the current vanishes the parameter $\eta$, and thus the relative magnitude $t_{\rm so}/t_s$, follows via $\eta = (\phi_R-\phi_L)/2$, see Fig.~\ref{fig:plot2}(b).
We note here that there is no requirement on the actual magnitude of the difference $|B_L-B_R|$: As long as the line shape of the resulting leakage current can be detected, the stopping points can be located.
Also if the fields are equal in magnitude but there still is a sizable leakage current due to, e.g., spin relaxation processes, then the stopping points related to $\bs t_{\rm so}$ are still detectable by locating the minima of the current.

In the above we assumed accurate control over the two Zeeman fields $\bs B_{L,R}$ separately.
In many systems, however, especially in devices based on III-V materials such as GaAs and InAs, but also in some Si- and Ge-based systems, atoms that carry finite nuclear spin yield small quasistatic, but random effective magnetic fields acting on the localized spins, sometimes of the order of a few mT when there is a significant fraction of spinful nuclei.
This means that the total Zeeman fields $\bs B_{L,R} = \bs B^{\rm ext}_{L,R} + \bs K_{L,R}$ are the sum of the externally applied fields $\bs B^\text{ext}_{L,R}$ and random components $\bs K_{L,R}$ that cannot be controlled.

However, since only the direction of the two total Zeeman fields matters for the procedure described above, the effect of the random contribution from the nuclear fields can be suppressed simply by working in the large-field limit $B^\text{ext}_{L,R}\gg K$, where $K$ is the typical magnitude of the nuclear fields on the dots; residual details depending on the specific configuration of $\bs K_{L,R}$ will average out in a typical experiment, where the total measurement time exceeds the correlation time of the nuclear fields.

\begin{figure}[t]
	\centering
	\includegraphics[width=\linewidth]{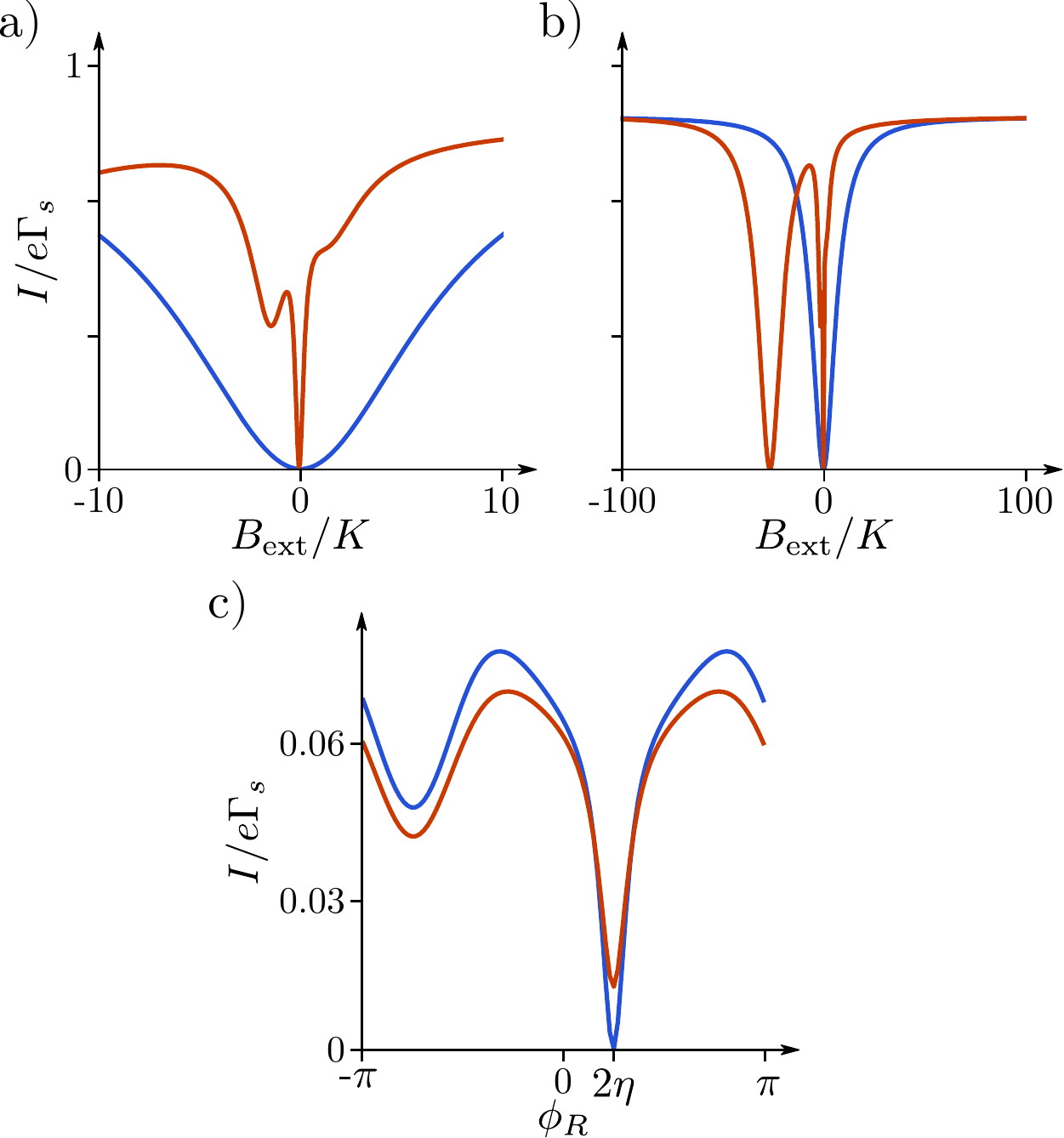} 
	\caption{
	(a,b) Calculated current as a function of the magnitude of a uniformly applied external field $B_\text{ext}$, assuming two different $g$-tensors on the two dots. In this plot we used $\bs B_L = \{0.76,0.32,0.34\}B_\text{ext}$ and $\bs B_R = \{1,0,0\}B_\text{ext}$ with $\Gamma_s = 0.1\ \mu$eV. 
	The blue lines show the case with no nuclear spins present, and the red lines show how adding two small random nuclear fields $\bs K_{L,R}$, drawn from a normal distribution with an r.m.s.\ value of $0.1~\mu$eV, drastically changes the behaviour of the current at small fields. 
	(c) Current as a function of $\phi_R$ with $B_L = 0.9$, $B_R=1$, $\theta_L = \theta_R = 3\pi/8$, and $\phi_L=0$. In the absence of nuclear fields (blue line) the current vanishes when the relative azimuthal angle $\delta\phi$ of two fields of different magnitude is equal to $2\eta$. 
	Averaging the current over random nuclear fields (red line) with the same distribution as used in (a,b), the current still has its minimum at $\delta \phi = 2\eta$.
	}
	\label{fig:plot3}
\end{figure}

We illustrate this in Fig.~\ref{fig:plot3}.
First, in Fig.~\ref{fig:plot3}(a,b) we exemplify the effect of one single static configuration of $\bs K_{L,R}$ on the leakage current:
The blue lines show the current as given by Eq.~\eqref{eq:current}, as a function of a uniformly applied magnetic field $B_\text{ext}$, in the absence of nuclear fields but assuming different $g$-tensors on the two dots (see the caption for the details).
For the red lines we added two randomly oriented nuclear fields with magnitudes drawn from a normal distribution with $\langle K_{L,R}^2 \rangle^{1/2} \equiv K = 0.1~\mu$eV.
We see that the difference is substantial at small fields, but vanishes at larger applied field.
In Fig.~\ref{fig:plot3}(c) we assume two external fields with $B_L = 0.9$, $B_R=1$ and $\theta_L = \theta_R = 3\pi/8$, looking for the current minimum as a function of their relative angle $\delta\phi$, in the absence of nuclear fields (blue line) and after averaging over many (finite) nuclear field configurations (red line).
Fig.~\ref{fig:plot3}(c) confirms that the averaging removes all sharp features, allowing again to locate the minimum in the current that is related to spin--orbit coupling, in the same way as in the case without nuclear fields.

\subsection{Homogeneous external field:\\ Hole-spin qubits with strong transverse confinement}
\label{sec:homogeneousfield}

Finally, we turn our attention to the more common situation where one can only control a homogeneous external field, yielding more or less equal Zeeman fields on the two dots.
Since the situation with $B_L=B_R$ corresponds to one of the stopping configurations discussed above, in this case finite nuclear fields are in fact required for obtaining a finite leakage current (in the absence of other spin relaxation processes).
Eq.~\eqref{eq:current} thus has to be averaged over the random fields $\bs K_{L,R}$ to find the leakage current that would be measured in a typical experiment, which is in general hard to do analytically.

One case, however, that can be treated analytically is potentially relevant for hole-based transport in quantum dots hosted in a quasi-two-dimensional carrier gas.
The valence band of most semiconductors is of $p$-type, which adds another threefold orbital angular momentum degree of freedom to the hole states.
Spin--orbit coupling splits off the states with total (orbital and spin) angular momentum $J=\frac{1}{2}$, leaving a four-dimensional $J=\frac{3}{2}$ low-energy subspace.
Out-of-plane confinement, used to create a two-dimensional hole gas, results in further splitting inside this subspace, lowering the energy of the so-called heavy holes (HHs) with $J_z = \pm \frac{3}{2}$ relative to the light holes (LHs) with $J_z = \pm \frac{1}{2}$.
For strong confinement this HH--LH splitting can become significant, in which case the low-energy confined states on the dots will mostly have a HH character.
Due to the $\pm\frac{3}{2}$ angular momentum carried by the two basis states, these states are to lowest order not expected to be coupled directly by the in-plane angular momentum operators $J^\pm$.
This is the reason why in the absence of significant HH--LH mixing most spin-dependent phenomena are usually highly anisotropic in the HH subspace:
The in-plane $g$-factor can be up to an order of magnitude smaller than the out-of-plane one~\cite{Qvist2022,Liles2021Dec,watzinger2016,Brauns2016,Bogan2017,Lu2017,Gradl2018,hofmann2019ArXiV,Miller2021}, hyperfine interaction with the residual nuclear spins could become effectively almost purely Ising-like~\cite{Fischer2008,Testelin2009,Prechtel2016,Bosco2021b} (although some experiments suggest that a significant $d$-shell state admixture can result in a much less anisotropic coupling than naively expected~\cite{Chekhovich2013,Machnikowski2019}), and also spin--orbit coupling inside the HH subspace will in general be more efficient along $J^z$.

\begin{figure}[t]
	\centering
	\includegraphics[width=\linewidth]{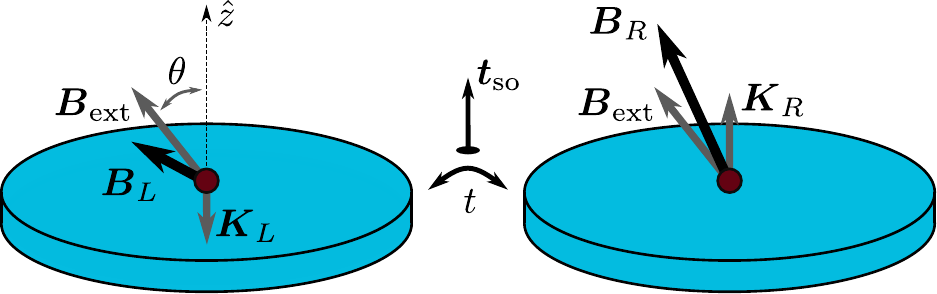} 
	\caption{Illustration of the orientation of the fields used for the analytical derivation in Sec.~\ref{sec:homogeneousfield}: The spin-orbit vector $\bs t_\text{so}$ and both nuclear fields $\bs K_{L,R}$ are assumed to be pointing along $\hat{z}$.
	The external Zeeman field $\bs B_\text{ext}$ is equal on the two dots, but can point in any direction.}
	\label{fig:figure_2}
\end{figure}
In this highly anisotropic limit we can thus assume that (i) the two nuclear fields are purely out-of-plane and (ii) the spin--orbit vector $\bs t_{\rm so}$ is also most likely to be out-of-plane.
In that case, the current (\ref{eq:current}) becomes a function of the fields $\bs B_{L,R} = \bs B_\text{ext} + K_{L,R}^z\hat z$, as illustrated in Fig.~\ref{fig:figure_2}.
The experimentally measured current then follows from averaging Eq.~(\ref{eq:current}) over $K_{L,R}^z$,
\begin{equation}
    \mathcal{I}_\text{av} = \int dK^z_L dK^z_R\, \frac{e^{-[(K_L^z)^2+(K_R^z)^2]/2K^2}}{4\pi K^2}\,I(\bs B_L,\bs B_R),
\end{equation}
where we have assumed the nuclear-field distributions to be Gaussian with mean zero and variance $K^2$.
Signatures of the hyperfine interaction that survive this averaging are again expected to be most prominent at small fields, where $B_\text{ext} \lesssim K$.
We will thus focus on the small-field limit, $\Gamma_s\gg K, B_\text{ext}$, where we find the approximate analytic result 
\begin{align}
    \frac{\mathcal{I}_\text{av}\Gamma_s}{eK^2} = {}&{} 2f\left(\alpha+ib^z\right)\left\{ 1+6f\left(\tfrac{1}{2}\beta\right)\beta^2 \right\} 
    \nonumber\\&\indent
    - f\left(\tfrac{1}{2}\alpha+ib^z\right)\left\{ 2+3f\left(\tfrac{1}{2}\beta\right)\beta^2 \right\},
    \label{eq:current_int}
\end{align}
where we have used the function
\begin{equation}
    f(x) = \frac{\sqrt{\pi}}{3}\ \mathrm{Re}\left\{x\right\} \mathrm{Re}\left\{e^{x^2} \mathrm{erfc}\left( x \right)\right\} - \frac{1}{3},
\end{equation}
with $\text{erfc}(x)$ being the complementary error function.
Furthermore, we introduced $\alpha = b^\parallel \cos\eta$ and $\beta = b^\parallel \sin\eta$ where $b^z = B^z_\text{ext}/K$ and $b^\parallel = \sqrt{(B_\text{ext}^x)^2+(B_\text{ext}^y)^2}/K$ give the out-of-plane and in-plane component of the external Zeeman field, respectively, in units of $K$.

\begin{figure}[t]
	\centering
	\includegraphics[width=\linewidth]{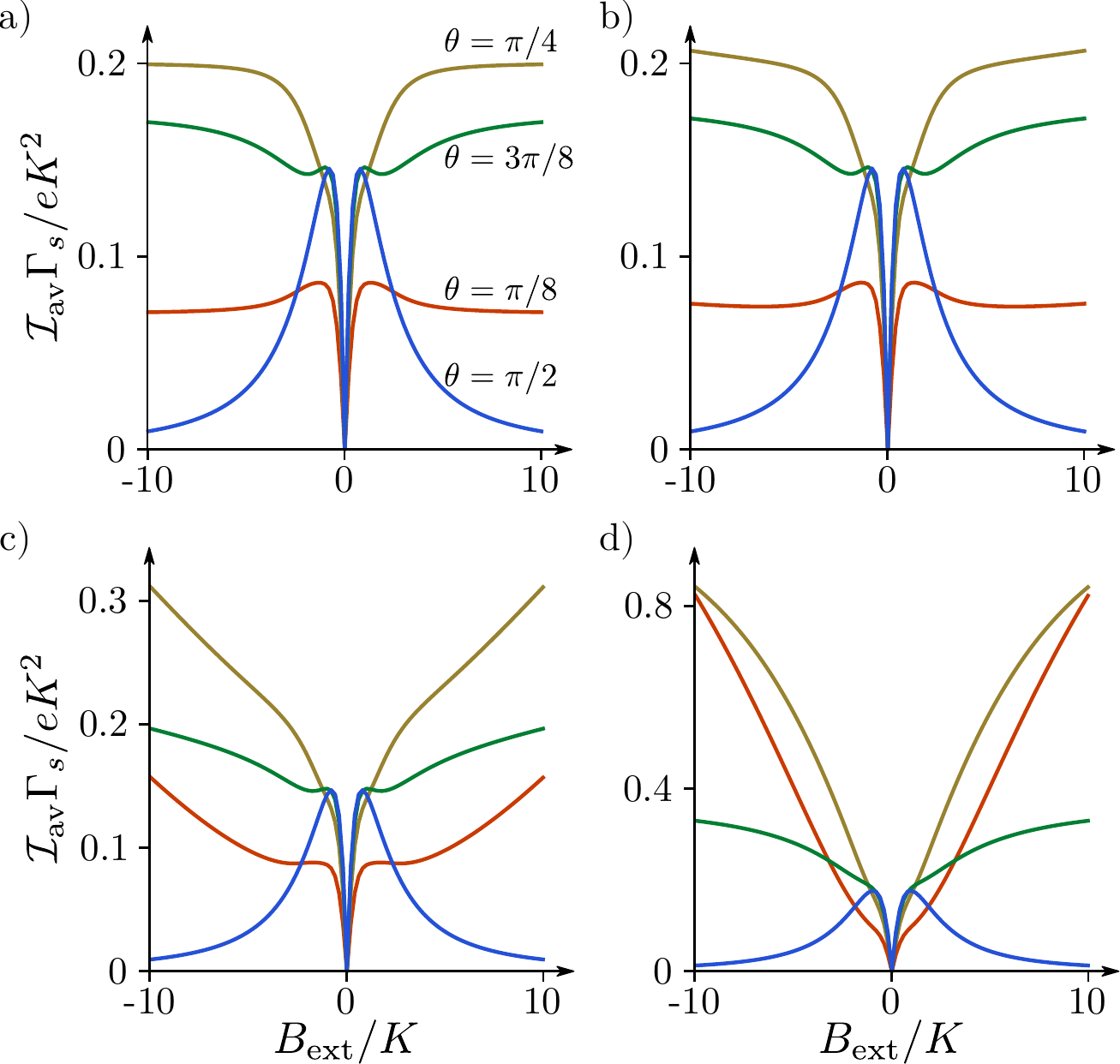} 
	\caption{
	The current as given by Eq.~\eqref{eq:current_int}, as a function of the magnitude of the applied field $\bs B_\text{ext}$. The four plots have an increasing magnitude of spin--orbit interaction: (a) $\eta=0$, (b) $\eta=0.02$, (c) $\eta=0.1$ and (d) $\eta=0.5$.
	In each plot the four traces correspond to four different orientations of the applied field, the corresponding polar angles of $\bs B_\text{ext}$ are indicated in (a) (same colors represent same orientations in all plots).
	}
	\label{fig:plots}
\end{figure}
In Fig.~\ref{fig:plots} we plot the current given by Eq.~\eqref{eq:current_int} as a function of the magnitude of the external magnetic field, for different orientations of the field and different strengths of spin–orbit coupling.
The four plots (a--d) show the current for different magnitudes of spin--orbit coupling ($\eta=0$, $\eta=0.02$, $\eta=0.1$ and $\eta=0.5$, respectively) and each plot contains four traces that assume a different orientation of $\bs B_\text{ext}$, the angle $\theta$ being the polar angle of the applied field (see Fig.~\ref{fig:figure_2}).
In all plots we used the parameters $\Gamma_s=15~\mu$eV and $K = 0.1~\mu$eV.

For all values of $\eta$ the current vanishes when the Zeeman field points along $\hat z$ (i.e., is parallel to $\bs K_{L,R}$ and $\bs t_\text{so}$) and at the point where $B_\text{ext}=0$, both of which are cases of the third stopping point mentioned above.

The spin--orbit-free case $\eta=0$ is shown in Fig.~\ref{fig:plots}(a). 
For most orientations of $\bs B_\text{ext}$, but most prominently for an in-plane field, we observe a peak in the current around zero field, with a width $\sim K$, that is split into a double peak by the stopping point at $B_\text{ext}=0$.
For large fields the current converges towards a direction-dependent limiting value $\mathcal{I}_\text{av}^\infty \approx 2 (e/\Gamma_s) K^2 \sin^2\theta /(4+\tan^2\theta)$.
This large-field current vanishes for $\theta=0$ (see above), but also for $\theta=\pi/2$, where the nuclear fields do not affect the magnitude of the total fields to leading order, resulting effectively in a blockade due to stopping point (i).

As illustrated in Fig.~\ref{fig:plots}(b--d), adding a finite spin--orbit coupling changes the current profiles:
On top of the narrow current peaks caused by the nuclear fields, we observe in most cases the characteristic spin--orbit-induced low-field current dip, the shape and width of which depend on $\eta$ and the direction of the applied field.
The large-field limiting current is typically larger than in the case of $\eta=0$, due to the efficient spin--orbit-induced spin mixing, which becomes more effective at larger fields.

Comparing Eq.~\ref{eq:current_int} with the experimentally measured low-field leakage current could thus give insight in the typical magnitude of the effective nuclear fields in the system as well as the total strength of the effective spin--orbit field, for the case of hole-based transport in systems with strong transverse confinement.
We note here that features similar to some observed in Fig.~\ref{fig:plots} (such as a low-field split peak on the background of a wider zero-field dip in the current) are indeed sometimes observed in such systems~\cite{Wang2018Aug,Marx2020Mar}.


\section{Conclusion}
Spin--orbit interaction is an important ingredient in low-dimensional semiconductor and hybrid structures, and understanding the detailed manifestation of the interaction is therefore essential. 
One of the mechanisms that couples the spin dynamics of localized carriers to the more easily detectable charge dynamics is the Pauli spin blockade that can occur in multi-quantum-dot structures.

In this paper we investigated in detail how the leakage current of a double quantum dot in spin blockade is affected by spin--orbit interaction. 
The main effect of spin--orbit interaction in such a situation is that it yields effectively a non-spin-conserving interdot tunnel coupling.
Using a simple few-level model Hamiltonian to describe the coupled spin-charge dynamics in the system, we derived a relatively compact analytic expression describing the leakage current through the blockade, including the detailed effect of the spin--orbit coupling.
From this result we could identify different so-called stopping points, for which the current vanishes, which allowed us to connect qualitative features in the current to both the magnitude and orientation of the effective spin--orbit field acting on the tunneling carriers.
This connection could thus provide a tool for characterizing the relevant spin--orbit parameters in multi-quantum-dot devices.

We then investigated the leakage current in more detail in the presence of randomly fluctuating nuclear spin baths that can couple to the localized carriers.
For the case of hole spins in a strongly confined two-dimensional hole gas, we derived an analytic expression for the low-field leakage current that includes averaging over the random effective nuclear field configurations on the two dots.
Comparing these results with the experimentally measured leakage current at small fields could provide additional information about the details of both the hyperfine and spin--orbit coupling in a system.

\section*{Acknowledgments}
This work is part of FRIPRO-Project No.~274853, which is funded by the Research Council of Norway (RCN), and was also partly supported by the Centers of Excellence funding scheme of the RCN, Project No.~262633, QuSpin.

\appendix

\section{Derivation of the current}
\label{app:mastser_eq}

We write the Hamiltonian \eqref{eq:hamtot} in the rotated basis $\ket{\tilde \psi} = e^{i\frac{\eta}{2}(\sigma_L^z - \sigma_R^z)}\ket{\psi}$ of one bright and three dark states, as explained in the main text.
Then, we introduce the (anti)symmetric magnetic fields $\bs B_\pm = \frac{1}{2}(\bs B_L \pm \bs B_R)$ and we define the auxiliary fields $\bs E_\pm = \{ B_\pm^x \cos \eta - B_\mp^y \sin \eta, B_\pm^y \cos \eta + B_\mp^x \sin \eta, B_\pm^z\}$ that incorporate the $\eta$-dependence of the Zeeman Hamiltonian.
In terms of these new fields, we can write the Zeeman terms as
\begin{align}
    H_B = {} & {} \frac{1}{\sqrt 2}\sum_\pm\Big[ (E_+^x\pm iE_+^y) \ket{\tilde T_0}\bra{\tilde T_\pm} \nonumber\\
    {} & {} \hspace{5em} + (\mp E_-^x-iE_-^y) \ket{\tilde S}\bra{\tilde T_\pm} + \text{H.c.} \Big] \nonumber\\
    {} & {} + E_+^z \big\{ \ket{\tilde T_+}\bra{\tilde T_+} - \ket{\tilde T_-}\bra{\tilde T_-} \big\} \nonumber\\
    {} & {} + E_-^z \big\{ \ket{\tilde S}\bra{\tilde T_0} + \ket{\tilde T_0}\bra{\tilde S} \big\},
\end{align}
which has exactly the same form as the usual (1,1) Zeeman Hamiltonian (\ref{eq:zeeman}) when written in a singlet-triplet basis \cite{jouravlev2006}, under the substitution $\bs B_\pm \to \bs E_\pm$.

The 3$\times$3 block of the Hamiltonian governing the subspace $\{\ket{\tilde T_+},\ket{\tilde T_0},\ket{\tilde T_-}\}$ thus describes a spin-1 system coupled to the spin--orbit-rotated effective field $\bs E_+$.
Applying the appropriate spin-1 rotation $\exp(i\alpha\bs J\cdot\hat{\bs n})$ (where $\bs J$ is the vector of spin-1 matrices and $\hat{\bs n}$ is the unit vector of rotation), we can diagonalize this block such that the full five-level Hamiltonian becomes
\begin{equation}
    H
    = \left(\begin{array}{ccccc}
        E_+ & 0 & 0 & 
        c & 0
        \\
        0 & 0 & 0 & 
        -d & 0 
        \\
        0 & 0 & -E_+ & 
        -c & 0 
        \\
        c & 
        -d & 
        -c & 0 & t
        \\
        0 & 0 & 0 & t & -\delta
    \end{array}\right),
    \label{eq:ham_end}
\end{equation}
where $t=\sqrt{t_s^2+t_\text{so}^2}$, and the (real) couplings between the triplets and the bright state $\ket{\tilde S}$ read
\begin{align}
    c = {} & {} \frac{E_-}{\sqrt{2}}\big\{ \left[\cos\theta_+\sin\theta_-\cos(\phi_+-\phi_-)-\cos\theta_-\sin\theta_+\right]^2 
    \nonumber\\
    {} & {} \hspace{.8cm} + \sin^2\theta_-\sin^2(\phi_+-\phi_-)\big\}^{1/2},
    \\
    d = {} & {}  E_-\left[\cos\theta_- \cos\theta_++\sin\theta_-\sin\theta_+\cos(\phi_+-\phi_-)\right],
\end{align}
with 
\begin{equation}
    \theta_\pm = \text{arccos}\left[\frac{E_\pm^z}{E_\pm}\right],
    \indent
    \phi_\pm = \text{arg}\left[E_\pm^x+iE_\pm^y\right],
\end{equation}
being the angles that define the orientation of the fields $\bs E_\pm$.
Having the Hamiltonian on this form is advantageous when calculating the current since it reduces the number of independent parameters from eight to five.

To obtain an analytical expression for the current through the system we then solve the master equation in steady state
\begin{equation}
    \frac{\partial \hat \rho}{\partial t} = -i[H,\hat \rho] + \bs\Gamma (\hat \rho) = 0,
    \label{eq:master_eq}
\end{equation}
where $\hat \rho$ is the five-level density matrix and $\bs \Gamma (\hat \rho) = -\frac{1}{2}\Gamma\{\hat{P}_{02},\hat{\rho}\} + \frac{1}{4}\Gamma (\mathbb{1}-\hat{P}_{02})\hat{\rho}_{02,02}$ the superoperator describing the fast tunneling processes to and from the reservoirs.
Here, $\Gamma$ is the characteristic rate of decay of $\ket{S_{02}}$ and subsequent reloading of one of the (1,1) states, and $\hat P_{02}=\ket{S_{02}}\bra{S_{02}}$ is the projector onto the state $\ket{S_{02}}$. 
After solving Eq.~\eqref{eq:master_eq} for the steady-state density matrix $\hat{\rho}^{\rm ss}$, the current through the double dot can be calculated from the expression $I = e\Gamma\hat \rho_{02,02}^{\rm ss}$, giving 
\begin{widetext}
\begin{align}
    \frac{8e\Gamma t^2c^2{d}^2E_+^2}{I} = \ & 4c^4{d}^2(4E_+^2+\Gamma^2+4\delta^2)\nonumber \\ &  + {d}^2[4E_+^6+4E_+^2t^4+d^4(4E_+^2+\Gamma^2+4\delta^2) -2{d}^2E_+^2(4E_+^2-4t^2+\Gamma^2+4\delta^2)]\nonumber \\
    & + 2{c}^2[E_+^2t^4+2d^4(4E_+^2+\Gamma^2+4\delta^2) +2{d}^2E_+^2(4E_+^2+2t^2+\Gamma^2+4\delta^2)].
    \label{eq:app_master_solution}
\end{align}
\end{widetext}
Eq.~\eqref{eq:current} in the main text then follows from assuming that we are in the strong-coupling regime, where $\Gamma\gg\delta,B_\pm$, and introducing the rate $\Gamma_s \equiv t^2/\Gamma$, which sets the scale of the effective decay of the (1,1) states.


\begin{thebibliography}{72}%
\makeatletter
\providecommand \@ifxundefined [1]{%
 \@ifx{#1\undefined}
}%
\providecommand \@ifnum [1]{%
 \ifnum #1\expandafter \@firstoftwo
 \else \expandafter \@secondoftwo
 \fi
}%
\providecommand \@ifx [1]{%
 \ifx #1\expandafter \@firstoftwo
 \else \expandafter \@secondoftwo
 \fi
}%
\providecommand \natexlab [1]{#1}%
\providecommand \enquote  [1]{``#1''}%
\providecommand \bibnamefont  [1]{#1}%
\providecommand \bibfnamefont [1]{#1}%
\providecommand \citenamefont [1]{#1}%
\providecommand \href@noop [0]{\@secondoftwo}%
\providecommand \href [0]{\begingroup \@sanitize@url \@href}%
\providecommand \@href[1]{\@@startlink{#1}\@@href}%
\providecommand \@@href[1]{\endgroup#1\@@endlink}%
\providecommand \@sanitize@url [0]{\catcode `\\12\catcode `\$12\catcode
  `\&12\catcode `\#12\catcode `\^12\catcode `\_12\catcode `\%12\relax}%
\providecommand \@@startlink[1]{}%
\providecommand \@@endlink[0]{}%
\providecommand \url  [0]{\begingroup\@sanitize@url \@url }%
\providecommand \@url [1]{\endgroup\@href {#1}{\urlprefix }}%
\providecommand \urlprefix  [0]{URL }%
\providecommand \Eprint [0]{\href }%
\providecommand \doibase [0]{http://dx.doi.org/}%
\providecommand \selectlanguage [0]{\@gobble}%
\providecommand \bibinfo  [0]{\@secondoftwo}%
\providecommand \bibfield  [0]{\@secondoftwo}%
\providecommand \translation [1]{[#1]}%
\providecommand \BibitemOpen [0]{}%
\providecommand \bibitemStop [0]{}%
\providecommand \bibitemNoStop [0]{.\EOS\space}%
\providecommand \EOS [0]{\spacefactor3000\relax}%
\providecommand \BibitemShut  [1]{\csname bibitem#1\endcsname}%
\let\auto@bib@innerbib\@empty
\bibitem [{\citenamefont {Sau}\ \emph {et~al.}(2010)\citenamefont {Sau},
  \citenamefont {Lutchyn}, \citenamefont {Tewari},\ and\ \citenamefont
  {Das~Sarma}}]{sau2010generic}%
  \BibitemOpen
  \bibfield  {author} {\bibinfo {author} {\bibfnamefont {J.~D.}\ \bibnamefont
  {Sau}}, \bibinfo {author} {\bibfnamefont {R.~M.}\ \bibnamefont {Lutchyn}},
  \bibinfo {author} {\bibfnamefont {S.}~\bibnamefont {Tewari}}, \ and\ \bibinfo
  {author} {\bibfnamefont {S.}~\bibnamefont {Das~Sarma}},\ }\href {\doibase
  10.1103/PhysRevLett.104.040502} {\bibfield  {journal} {\bibinfo  {journal}
  {Phys. Rev. Lett.}\ }\textbf {\bibinfo {volume} {104}},\ \bibinfo {pages}
  {040502} (\bibinfo {year} {2010})}\BibitemShut {NoStop}%
\bibitem [{\citenamefont {Lutchyn}\ \emph {et~al.}(2010)\citenamefont
  {Lutchyn}, \citenamefont {Sau},\ and\ \citenamefont
  {Das~Sarma}}]{lutchyn2010majorana}%
  \BibitemOpen
  \bibfield  {author} {\bibinfo {author} {\bibfnamefont {R.~M.}\ \bibnamefont
  {Lutchyn}}, \bibinfo {author} {\bibfnamefont {J.~D.}\ \bibnamefont {Sau}}, \
  and\ \bibinfo {author} {\bibfnamefont {S.}~\bibnamefont {Das~Sarma}},\ }\href
  {\doibase 10.1103/PhysRevLett.105.077001} {\bibfield  {journal} {\bibinfo
  {journal} {Phys. Rev. Lett.}\ }\textbf {\bibinfo {volume} {105}},\ \bibinfo
  {pages} {077001} (\bibinfo {year} {2010})}\BibitemShut {NoStop}%
\bibitem [{\citenamefont {Oreg}\ \emph {et~al.}(2010)\citenamefont {Oreg},
  \citenamefont {Refael},\ and\ \citenamefont {von Oppen}}]{oreg2010helical}%
  \BibitemOpen
  \bibfield  {author} {\bibinfo {author} {\bibfnamefont {Y.}~\bibnamefont
  {Oreg}}, \bibinfo {author} {\bibfnamefont {G.}~\bibnamefont {Refael}}, \ and\
  \bibinfo {author} {\bibfnamefont {F.}~\bibnamefont {von Oppen}},\ }\href
  {\doibase 10.1103/PhysRevLett.105.177002} {\bibfield  {journal} {\bibinfo
  {journal} {Phys. Rev. Lett.}\ }\textbf {\bibinfo {volume} {105}},\ \bibinfo
  {pages} {177002} (\bibinfo {year} {2010})}\BibitemShut {NoStop}%
\bibitem [{\citenamefont {Leijnse}\ and\ \citenamefont
  {Flensberg}(2012)}]{leijnse2012introduction}%
  \BibitemOpen
  \bibfield  {author} {\bibinfo {author} {\bibfnamefont {M.}~\bibnamefont
  {Leijnse}}\ and\ \bibinfo {author} {\bibfnamefont {K.}~\bibnamefont
  {Flensberg}},\ }\href {\doibase 10.1088/0268-1242/27/12/124003} {\bibfield
  {journal} {\bibinfo  {journal} {Semicond. Science and Techn.}\ }\textbf
  {\bibinfo {volume} {27}},\ \bibinfo {pages} {124003} (\bibinfo {year}
  {2012})}\BibitemShut {NoStop}%
\bibitem [{\citenamefont {Beenakker}(2013)}]{beenakker2013search}%
  \BibitemOpen
  \bibfield  {author} {\bibinfo {author} {\bibfnamefont {C.}~\bibnamefont
  {Beenakker}},\ }\href {\doibase 10.1146/annurev-conmatphys-030212-184337}
  {\bibfield  {journal} {\bibinfo  {journal} {Ann. Rev.of Cond. Matt. Phys.}\
  }\textbf {\bibinfo {volume} {4}},\ \bibinfo {pages} {113} (\bibinfo {year}
  {2013})}\BibitemShut {NoStop}%
\bibitem [{\citenamefont {Lutchyn}\ \emph {et~al.}(2018)\citenamefont
  {Lutchyn}, \citenamefont {Bakkers}, \citenamefont {Kouwenhoven},
  \citenamefont {Krogstrup}, \citenamefont {Marcus},\ and\ \citenamefont
  {Oreg}}]{Lutchyn2018May}%
  \BibitemOpen
  \bibfield  {author} {\bibinfo {author} {\bibfnamefont {R.~M.}\ \bibnamefont
  {Lutchyn}}, \bibinfo {author} {\bibfnamefont {E.~P. A.~M.}\ \bibnamefont
  {Bakkers}}, \bibinfo {author} {\bibfnamefont {L.~P.}\ \bibnamefont
  {Kouwenhoven}}, \bibinfo {author} {\bibfnamefont {P.}~\bibnamefont
  {Krogstrup}}, \bibinfo {author} {\bibfnamefont {C.~M.}\ \bibnamefont
  {Marcus}}, \ and\ \bibinfo {author} {\bibfnamefont {Y.}~\bibnamefont
  {Oreg}},\ }\href {\doibase 10.1038/s41578-018-0003-1} {\bibfield  {journal}
  {\bibinfo  {journal} {Nat. Rev. Mater.}\ }\textbf {\bibinfo {volume} {3}},\
  \bibinfo {pages} {52} (\bibinfo {year} {2018})}\BibitemShut {NoStop}%
\bibitem [{\citenamefont {Kitaev}(2003)}]{kitaev2003fault}%
  \BibitemOpen
  \bibfield  {author} {\bibinfo {author} {\bibfnamefont {A.~Y.}\ \bibnamefont
  {Kitaev}},\ }\href
  {https://www.sciencedirect.com/science/article/pii/S0003491602000180}
  {\bibfield  {journal} {\bibinfo  {journal} {Ann. Phys.}\ }\textbf {\bibinfo
  {volume} {303}},\ \bibinfo {pages} {2} (\bibinfo {year} {2003})}\BibitemShut
  {NoStop}%
\bibitem [{\citenamefont {Nayak}\ \emph {et~al.}(2008)\citenamefont {Nayak},
  \citenamefont {Simon}, \citenamefont {Stern}, \citenamefont {Freedman},\ and\
  \citenamefont {Das~Sarma}}]{nayak2008non}%
  \BibitemOpen
  \bibfield  {author} {\bibinfo {author} {\bibfnamefont {C.}~\bibnamefont
  {Nayak}}, \bibinfo {author} {\bibfnamefont {S.~H.}\ \bibnamefont {Simon}},
  \bibinfo {author} {\bibfnamefont {A.}~\bibnamefont {Stern}}, \bibinfo
  {author} {\bibfnamefont {M.}~\bibnamefont {Freedman}}, \ and\ \bibinfo
  {author} {\bibfnamefont {S.}~\bibnamefont {Das~Sarma}},\ }\href {\doibase
  10.1103/RevModPhys.80.1083} {\bibfield  {journal} {\bibinfo  {journal} {Rev.
  Mod. Phys.}\ }\textbf {\bibinfo {volume} {80}},\ \bibinfo {pages} {1083}
  (\bibinfo {year} {2008})}\BibitemShut {NoStop}%
\bibitem [{\citenamefont {{Das Sarma}}\ \emph {et~al.}(2015)\citenamefont {{Das
  Sarma}}, \citenamefont {Freedman},\ and\ \citenamefont
  {Nayak}}]{sarma2015majorana}%
  \BibitemOpen
  \bibfield  {author} {\bibinfo {author} {\bibfnamefont {S.}~\bibnamefont {{Das
  Sarma}}}, \bibinfo {author} {\bibfnamefont {M.}~\bibnamefont {Freedman}}, \
  and\ \bibinfo {author} {\bibfnamefont {C.}~\bibnamefont {Nayak}},\ }\href
  {https://www.nature.com/articles/npjqi20151} {\bibfield  {journal} {\bibinfo
  {journal} {npj Quant. Inf.}\ }\textbf {\bibinfo {volume} {1}},\ \bibinfo
  {pages} {1} (\bibinfo {year} {2015})}\BibitemShut {NoStop}%
\bibitem [{\citenamefont {Stanescu}(2016)}]{stanescu2016introduction}%
  \BibitemOpen
  \bibfield  {author} {\bibinfo {author} {\bibfnamefont {T.~D.}\ \bibnamefont
  {Stanescu}},\ }\href@noop {} {\emph {\bibinfo {title} {{Introduction to
  Topological Quantum Matter \& Quantum Computation}}}}\ (\bibinfo  {publisher}
  {CRC Press},\ \bibinfo {year} {2016})\BibitemShut {NoStop}%
\bibitem [{\citenamefont {Petersson}\ \emph {et~al.}(2012)\citenamefont
  {Petersson}, \citenamefont {McFaul}, \citenamefont {Schroer}, \citenamefont
  {Jung}, \citenamefont {Taylor}, \citenamefont {Houck},\ and\ \citenamefont
  {Petta}}]{Petersson2012Oct}%
  \BibitemOpen
  \bibfield  {author} {\bibinfo {author} {\bibfnamefont {K.~D.}\ \bibnamefont
  {Petersson}}, \bibinfo {author} {\bibfnamefont {L.~W.}\ \bibnamefont
  {McFaul}}, \bibinfo {author} {\bibfnamefont {M.~D.}\ \bibnamefont {Schroer}},
  \bibinfo {author} {\bibfnamefont {M.}~\bibnamefont {Jung}}, \bibinfo {author}
  {\bibfnamefont {J.~M.}\ \bibnamefont {Taylor}}, \bibinfo {author}
  {\bibfnamefont {A.~A.}\ \bibnamefont {Houck}}, \ and\ \bibinfo {author}
  {\bibfnamefont {J.~R.}\ \bibnamefont {Petta}},\ }\href {\doibase
  10.1038/nature11559} {\bibfield  {journal} {\bibinfo  {journal} {Nature}\
  }\textbf {\bibinfo {volume} {490}},\ \bibinfo {pages} {380} (\bibinfo {year}
  {2012})}\BibitemShut {NoStop}%
\bibitem [{\citenamefont {Samkharadze}\ \emph {et~al.}(2018)\citenamefont
  {Samkharadze}, \citenamefont {Zheng}, \citenamefont {Kalhor}, \citenamefont
  {Brousse}, \citenamefont {Sammak}, \citenamefont {Mendes}, \citenamefont
  {Blais}, \citenamefont {Scappucci},\ and\ \citenamefont
  {Vandersypen}}]{Samkharadze2018}%
  \BibitemOpen
  \bibfield  {author} {\bibinfo {author} {\bibfnamefont {N.}~\bibnamefont
  {Samkharadze}}, \bibinfo {author} {\bibfnamefont {G.}~\bibnamefont {Zheng}},
  \bibinfo {author} {\bibfnamefont {N.}~\bibnamefont {Kalhor}}, \bibinfo
  {author} {\bibfnamefont {D.}~\bibnamefont {Brousse}}, \bibinfo {author}
  {\bibfnamefont {A.}~\bibnamefont {Sammak}}, \bibinfo {author} {\bibfnamefont
  {U.~C.}\ \bibnamefont {Mendes}}, \bibinfo {author} {\bibfnamefont
  {A.}~\bibnamefont {Blais}}, \bibinfo {author} {\bibfnamefont
  {G.}~\bibnamefont {Scappucci}}, \ and\ \bibinfo {author} {\bibfnamefont
  {L.~M.~K.}\ \bibnamefont {Vandersypen}},\ }\href {\doibase
  10.1126/science.aar4054} {\bibfield  {journal} {\bibinfo  {journal}
  {Science}\ }\textbf {\bibinfo {volume} {359}},\ \bibinfo {pages} {1123}
  (\bibinfo {year} {2018})}\BibitemShut {NoStop}%
\bibitem [{\citenamefont {Burkard}\ \emph {et~al.}(2020)\citenamefont
  {Burkard}, \citenamefont {Gullans}, \citenamefont {Mi},\ and\ \citenamefont
  {Petta}}]{burkard2020superconductor}%
  \BibitemOpen
  \bibfield  {author} {\bibinfo {author} {\bibfnamefont {G.}~\bibnamefont
  {Burkard}}, \bibinfo {author} {\bibfnamefont {M.~J.}\ \bibnamefont
  {Gullans}}, \bibinfo {author} {\bibfnamefont {X.}~\bibnamefont {Mi}}, \ and\
  \bibinfo {author} {\bibfnamefont {J.~R.}\ \bibnamefont {Petta}},\ }\href
  {https://www.nature.com/articles/s42254-019-0135-2} {\bibfield  {journal}
  {\bibinfo  {journal} {Nat. Rev. Phys.}\ }\textbf {\bibinfo {volume} {2}},\
  \bibinfo {pages} {129} (\bibinfo {year} {2020})}\BibitemShut {NoStop}%
\bibitem [{\citenamefont {Rashba}\ and\ \citenamefont
  {Efros}(2003)}]{Rashba2003Sep}%
  \BibitemOpen
  \bibfield  {author} {\bibinfo {author} {\bibfnamefont {E.~I.}\ \bibnamefont
  {Rashba}}\ and\ \bibinfo {author} {\bibfnamefont {{\relax Al}.~L.}\
  \bibnamefont {Efros}},\ }\href {\doibase 10.1103/PhysRevLett.91.126405}
  {\bibfield  {journal} {\bibinfo  {journal} {Phys. Rev. Lett.}\ }\textbf
  {\bibinfo {volume} {91}},\ \bibinfo {pages} {126405} (\bibinfo {year}
  {2003})}\BibitemShut {NoStop}%
\bibitem [{\citenamefont
  {{\ifmmode\check{Z}\else\v{Z}\fi}uti{\ifmmode\acute{c}\else\'{c}\fi}}\ \emph
  {et~al.}(2004)\citenamefont
  {{\ifmmode\check{Z}\else\v{Z}\fi}uti{\ifmmode\acute{c}\else\'{c}\fi}},
  \citenamefont {Fabian},\ and\ \citenamefont {Das~Sarma}}]{Zutic2004Apr}%
  \BibitemOpen
  \bibfield  {author} {\bibinfo {author} {\bibfnamefont {I.}~\bibnamefont
  {{\ifmmode\check{Z}\else\v{Z}\fi}uti{\ifmmode\acute{c}\else\'{c}\fi}}},
  \bibinfo {author} {\bibfnamefont {J.}~\bibnamefont {Fabian}}, \ and\ \bibinfo
  {author} {\bibfnamefont {S.}~\bibnamefont {Das~Sarma}},\ }\href {\doibase
  10.1103/RevModPhys.76.323} {\bibfield  {journal} {\bibinfo  {journal} {Rev.
  Mod. Phys.}\ }\textbf {\bibinfo {volume} {76}},\ \bibinfo {pages} {323}
  (\bibinfo {year} {2004})}\BibitemShut {NoStop}%
\bibitem [{\citenamefont {Nadj-Perge}\ \emph {et~al.}(2010)\citenamefont
  {Nadj-Perge}, \citenamefont {Frolov}, \citenamefont {Bakkers},\ and\
  \citenamefont {Kouwenhoven}}]{Nadj-Perge2010Dec}%
  \BibitemOpen
  \bibfield  {author} {\bibinfo {author} {\bibfnamefont {S.}~\bibnamefont
  {Nadj-Perge}}, \bibinfo {author} {\bibfnamefont {S.~M.}\ \bibnamefont
  {Frolov}}, \bibinfo {author} {\bibfnamefont {E.~P. A.~M.}\ \bibnamefont
  {Bakkers}}, \ and\ \bibinfo {author} {\bibfnamefont {L.~P.}\ \bibnamefont
  {Kouwenhoven}},\ }\href {\doibase 10.1038/nature09682} {\bibfield  {journal}
  {\bibinfo  {journal} {Nature}\ }\textbf {\bibinfo {volume} {468}},\ \bibinfo
  {pages} {1084} (\bibinfo {year} {2010})}\BibitemShut {NoStop}%
\bibitem [{\citenamefont {Loss}\ and\ \citenamefont
  {DiVincenzo}(1998)}]{Loss1998}%
  \BibitemOpen
  \bibfield  {author} {\bibinfo {author} {\bibfnamefont {D.}~\bibnamefont
  {Loss}}\ and\ \bibinfo {author} {\bibfnamefont {D.~P.}\ \bibnamefont
  {DiVincenzo}},\ }\href {\doibase 10.1103/physreva.57.120} {\bibfield
  {journal} {\bibinfo  {journal} {Phys. Rev. A}\ }\textbf {\bibinfo {volume}
  {57}},\ \bibinfo {pages} {120} (\bibinfo {year} {1998})}\BibitemShut
  {NoStop}%
\bibitem [{\citenamefont {Hanson}\ \emph {et~al.}(2007)\citenamefont {Hanson},
  \citenamefont {Kouwenhoven}, \citenamefont {Petta}, \citenamefont {Tarucha},\
  and\ \citenamefont {Vandersypen}}]{Hanson2007Oct}%
  \BibitemOpen
  \bibfield  {author} {\bibinfo {author} {\bibfnamefont {R.}~\bibnamefont
  {Hanson}}, \bibinfo {author} {\bibfnamefont {L.~P.}\ \bibnamefont
  {Kouwenhoven}}, \bibinfo {author} {\bibfnamefont {J.~R.}\ \bibnamefont
  {Petta}}, \bibinfo {author} {\bibfnamefont {S.}~\bibnamefont {Tarucha}}, \
  and\ \bibinfo {author} {\bibfnamefont {L.~M.~K.}\ \bibnamefont
  {Vandersypen}},\ }\href {\doibase 10.1103/RevModPhys.79.1217} {\bibfield
  {journal} {\bibinfo  {journal} {Rev. Mod. Phys.}\ }\textbf {\bibinfo {volume}
  {79}},\ \bibinfo {pages} {1217} (\bibinfo {year} {2007})}\BibitemShut
  {NoStop}%
\bibitem [{\citenamefont {Kloeffel}\ and\ \citenamefont
  {Loss}(2013)}]{Kloeffel2013Mar}%
  \BibitemOpen
  \bibfield  {author} {\bibinfo {author} {\bibfnamefont {C.}~\bibnamefont
  {Kloeffel}}\ and\ \bibinfo {author} {\bibfnamefont {D.}~\bibnamefont
  {Loss}},\ }\href {\doibase 10.1146/annurev-conmatphys-030212-184248}
  {\bibfield  {journal} {\bibinfo  {journal} {Annu. Rev. Condens. Matter
  Phys.}\ }\textbf {\bibinfo {volume} {4}},\ \bibinfo {pages} {51} (\bibinfo
  {year} {2013})}\BibitemShut {NoStop}%
\bibitem [{\citenamefont {Chatterjee}\ \emph {et~al.}(2021)\citenamefont
  {Chatterjee}, \citenamefont {Stevenson}, \citenamefont {De~Franceschi},
  \citenamefont {Morello}, \citenamefont {de~Leon},\ and\ \citenamefont
  {Kuemmeth}}]{Chatterjee2021Mar}%
  \BibitemOpen
  \bibfield  {author} {\bibinfo {author} {\bibfnamefont {A.}~\bibnamefont
  {Chatterjee}}, \bibinfo {author} {\bibfnamefont {P.}~\bibnamefont
  {Stevenson}}, \bibinfo {author} {\bibfnamefont {S.}~\bibnamefont
  {De~Franceschi}}, \bibinfo {author} {\bibfnamefont {A.}~\bibnamefont
  {Morello}}, \bibinfo {author} {\bibfnamefont {N.~P.}\ \bibnamefont
  {de~Leon}}, \ and\ \bibinfo {author} {\bibfnamefont {F.}~\bibnamefont
  {Kuemmeth}},\ }\href {\doibase 10.1038/s42254-021-00283-9} {\bibfield
  {journal} {\bibinfo  {journal} {Nat. Rev. Phys.}\ }\textbf {\bibinfo {volume}
  {3}},\ \bibinfo {pages} {157} (\bibinfo {year} {2021})}\BibitemShut {NoStop}%
\bibitem [{\citenamefont {Burkard}\ \emph {et~al.}(2021)\citenamefont
  {Burkard}, \citenamefont {Ladd}, \citenamefont {Nichol}, \citenamefont
  {Pan},\ and\ \citenamefont {Petta}}]{burkard2021semiconductor}%
  \BibitemOpen
  \bibfield  {author} {\bibinfo {author} {\bibfnamefont {G.}~\bibnamefont
  {Burkard}}, \bibinfo {author} {\bibfnamefont {T.~D.}\ \bibnamefont {Ladd}},
  \bibinfo {author} {\bibfnamefont {J.~M.}\ \bibnamefont {Nichol}}, \bibinfo
  {author} {\bibfnamefont {A.}~\bibnamefont {Pan}}, \ and\ \bibinfo {author}
  {\bibfnamefont {J.~R.}\ \bibnamefont {Petta}},\ }\href
  {https://arxiv.org/abs/2112.08863} {\bibfield  {journal} {\bibinfo  {journal}
  {arXiv:2112.08863}\ } (\bibinfo {year} {2021})}\BibitemShut {NoStop}%
\bibitem [{\citenamefont {Golovach}\ \emph {et~al.}(2006)\citenamefont
  {Golovach}, \citenamefont {Borhani},\ and\ \citenamefont
  {Loss}}]{Golovach2006Oct}%
  \BibitemOpen
  \bibfield  {author} {\bibinfo {author} {\bibfnamefont {V.~N.}\ \bibnamefont
  {Golovach}}, \bibinfo {author} {\bibfnamefont {M.}~\bibnamefont {Borhani}}, \
  and\ \bibinfo {author} {\bibfnamefont {D.}~\bibnamefont {Loss}},\ }\href
  {\doibase 10.1103/PhysRevB.74.165319} {\bibfield  {journal} {\bibinfo
  {journal} {Phys. Rev. B}\ }\textbf {\bibinfo {volume} {74}},\ \bibinfo
  {pages} {165319} (\bibinfo {year} {2006})}\BibitemShut {NoStop}%
\bibitem [{\citenamefont {Flindt}\ \emph {et~al.}(2006)\citenamefont {Flindt},
  \citenamefont {S{\o}rensen},\ and\ \citenamefont
  {Flensberg}}]{Flindt2006Dec}%
  \BibitemOpen
  \bibfield  {author} {\bibinfo {author} {\bibfnamefont {C.}~\bibnamefont
  {Flindt}}, \bibinfo {author} {\bibfnamefont {A.~S.}\ \bibnamefont
  {S{\o}rensen}}, \ and\ \bibinfo {author} {\bibfnamefont {K.}~\bibnamefont
  {Flensberg}},\ }\href {\doibase 10.1103/PhysRevLett.97.240501} {\bibfield
  {journal} {\bibinfo  {journal} {Phys. Rev. Lett.}\ }\textbf {\bibinfo
  {volume} {97}},\ \bibinfo {pages} {240501} (\bibinfo {year}
  {2006})}\BibitemShut {NoStop}%
\bibitem [{\citenamefont {Khaetskii}\ and\ \citenamefont
  {Nazarov}(2000)}]{Khaetskii2000}%
  \BibitemOpen
  \bibfield  {author} {\bibinfo {author} {\bibfnamefont {A.~V.}\ \bibnamefont
  {Khaetskii}}\ and\ \bibinfo {author} {\bibfnamefont {Y.~V.}\ \bibnamefont
  {Nazarov}},\ }\href {\doibase 10.1103/PhysRevB.61.12639} {\bibfield
  {journal} {\bibinfo  {journal} {Phys. Rev. B}\ }\textbf {\bibinfo {volume}
  {61}},\ \bibinfo {pages} {12639} (\bibinfo {year} {2000})}\BibitemShut
  {NoStop}%
\bibitem [{\citenamefont {Khaetskii}\ and\ \citenamefont
  {Nazarov}(2001)}]{Khaetskii2001}%
  \BibitemOpen
  \bibfield  {author} {\bibinfo {author} {\bibfnamefont {A.~V.}\ \bibnamefont
  {Khaetskii}}\ and\ \bibinfo {author} {\bibfnamefont {Y.~V.}\ \bibnamefont
  {Nazarov}},\ }\href {\doibase 10.1103/PhysRevB.64.125316} {\bibfield
  {journal} {\bibinfo  {journal} {Phys. Rev. B}\ }\textbf {\bibinfo {volume}
  {64}},\ \bibinfo {pages} {125316} (\bibinfo {year} {2001})}\BibitemShut
  {NoStop}%
\bibitem [{\citenamefont {Golovach}\ \emph {et~al.}(2004)\citenamefont
  {Golovach}, \citenamefont {Khaetskii},\ and\ \citenamefont
  {Loss}}]{golovach2004phonon}%
  \BibitemOpen
  \bibfield  {author} {\bibinfo {author} {\bibfnamefont {V.~N.}\ \bibnamefont
  {Golovach}}, \bibinfo {author} {\bibfnamefont {A.}~\bibnamefont {Khaetskii}},
  \ and\ \bibinfo {author} {\bibfnamefont {D.}~\bibnamefont {Loss}},\ }\href
  {\doibase 10.1103/PhysRevLett.93.016601} {\bibfield  {journal} {\bibinfo
  {journal} {Phys. Rev. Lett.}\ }\textbf {\bibinfo {volume} {93}},\ \bibinfo
  {pages} {016601} (\bibinfo {year} {2004})}\BibitemShut {NoStop}%
\bibitem [{\citenamefont {Maurand}\ \emph {et~al.}(2016)\citenamefont
  {Maurand}, \citenamefont {Jehl}, \citenamefont {Kotekar-Patil}, \citenamefont
  {Corna}, \citenamefont {Bohuslavskyi}, \citenamefont {Lavi{\'e}ville},
  \citenamefont {Hutin}, \citenamefont {Barraud}, \citenamefont {Vinet},
  \citenamefont {Sanquer},\ and\ \citenamefont {De~Franceschi}}]{Maurand2016}%
  \BibitemOpen
  \bibfield  {author} {\bibinfo {author} {\bibfnamefont {R.}~\bibnamefont
  {Maurand}}, \bibinfo {author} {\bibfnamefont {X.}~\bibnamefont {Jehl}},
  \bibinfo {author} {\bibfnamefont {D.}~\bibnamefont {Kotekar-Patil}}, \bibinfo
  {author} {\bibfnamefont {A.}~\bibnamefont {Corna}}, \bibinfo {author}
  {\bibfnamefont {H.}~\bibnamefont {Bohuslavskyi}}, \bibinfo {author}
  {\bibfnamefont {R.}~\bibnamefont {Lavi{\'e}ville}}, \bibinfo {author}
  {\bibfnamefont {L.}~\bibnamefont {Hutin}}, \bibinfo {author} {\bibfnamefont
  {S.}~\bibnamefont {Barraud}}, \bibinfo {author} {\bibfnamefont
  {M.}~\bibnamefont {Vinet}}, \bibinfo {author} {\bibfnamefont
  {M.}~\bibnamefont {Sanquer}}, \ and\ \bibinfo {author} {\bibfnamefont
  {S.}~\bibnamefont {De~Franceschi}},\ }\href {\doibase 10.1038/ncomms13575}
  {\bibfield  {journal} {\bibinfo  {journal} {Nat. Comm.}\ }\textbf {\bibinfo
  {volume} {7}},\ \bibinfo {pages} {13575} (\bibinfo {year}
  {2016})}\BibitemShut {NoStop}%
\bibitem [{\citenamefont {Watzinger}\ \emph {et~al.}(2018)\citenamefont
  {Watzinger}, \citenamefont {Kuku{\v{c}}ka}, \citenamefont
  {Vuku{\v{s}}i{\'{c}}}, \citenamefont {Gao}, \citenamefont {Wang},
  \citenamefont {Sch{\"a}ffler}, \citenamefont {Zhang},\ and\ \citenamefont
  {Katsaros}}]{Watzinger2018}%
  \BibitemOpen
  \bibfield  {author} {\bibinfo {author} {\bibfnamefont {H.}~\bibnamefont
  {Watzinger}}, \bibinfo {author} {\bibfnamefont {J.}~\bibnamefont
  {Kuku{\v{c}}ka}}, \bibinfo {author} {\bibfnamefont {L.}~\bibnamefont
  {Vuku{\v{s}}i{\'{c}}}}, \bibinfo {author} {\bibfnamefont {F.}~\bibnamefont
  {Gao}}, \bibinfo {author} {\bibfnamefont {T.}~\bibnamefont {Wang}}, \bibinfo
  {author} {\bibfnamefont {F.}~\bibnamefont {Sch{\"a}ffler}}, \bibinfo {author}
  {\bibfnamefont {J.-J.}\ \bibnamefont {Zhang}}, \ and\ \bibinfo {author}
  {\bibfnamefont {G.}~\bibnamefont {Katsaros}},\ }\href {\doibase
  10.1038/s41467-018-06418-4} {\bibfield  {journal} {\bibinfo  {journal} {Nat.
  Comm.}\ }\textbf {\bibinfo {volume} {9}},\ \bibinfo {pages} {3902} (\bibinfo
  {year} {2018})}\BibitemShut {NoStop}%
\bibitem [{\citenamefont {Vukušić}\ \emph {et~al.}(2018)\citenamefont
  {Vukušić}, \citenamefont {Kukučka}, \citenamefont {Watzinger},
  \citenamefont {Milem}, \citenamefont {Schäffler},\ and\ \citenamefont
  {Katsaros}}]{Vukusic2018}%
  \BibitemOpen
  \bibfield  {author} {\bibinfo {author} {\bibfnamefont {L.}~\bibnamefont
  {Vukušić}}, \bibinfo {author} {\bibfnamefont {J.}~\bibnamefont {Kukučka}},
  \bibinfo {author} {\bibfnamefont {H.}~\bibnamefont {Watzinger}}, \bibinfo
  {author} {\bibfnamefont {J.~M.}\ \bibnamefont {Milem}}, \bibinfo {author}
  {\bibfnamefont {F.}~\bibnamefont {Schäffler}}, \ and\ \bibinfo {author}
  {\bibfnamefont {G.}~\bibnamefont {Katsaros}},\ }\href {\doibase
  10.1021/acs.nanolett.8b03217} {\bibfield  {journal} {\bibinfo  {journal}
  {Nano Lett.}\ }\textbf {\bibinfo {volume} {18}},\ \bibinfo {pages} {7141}
  (\bibinfo {year} {2018})}\BibitemShut {NoStop}%
\bibitem [{\citenamefont {Crippa}\ \emph {et~al.}(2019)\citenamefont {Crippa},
  \citenamefont {Ezzouch}, \citenamefont {Apr{\'{a}}}, \citenamefont {Amisse},
  \citenamefont {Lavi{\'{e}}ville}, \citenamefont {Hutin}, \citenamefont
  {Bertrand}, \citenamefont {Vinet}, \citenamefont {Urdampilleta},
  \citenamefont {Meunier}, \citenamefont {Sanquer}, \citenamefont {Jehl},
  \citenamefont {Maurand},\ and\ \citenamefont {{De Franceschi}}}]{Crippa2019}%
  \BibitemOpen
  \bibfield  {author} {\bibinfo {author} {\bibfnamefont {A.}~\bibnamefont
  {Crippa}}, \bibinfo {author} {\bibfnamefont {R.}~\bibnamefont {Ezzouch}},
  \bibinfo {author} {\bibfnamefont {A.}~\bibnamefont {Apr{\'{a}}}}, \bibinfo
  {author} {\bibfnamefont {A.}~\bibnamefont {Amisse}}, \bibinfo {author}
  {\bibfnamefont {R.}~\bibnamefont {Lavi{\'{e}}ville}}, \bibinfo {author}
  {\bibfnamefont {L.}~\bibnamefont {Hutin}}, \bibinfo {author} {\bibfnamefont
  {B.}~\bibnamefont {Bertrand}}, \bibinfo {author} {\bibfnamefont
  {M.}~\bibnamefont {Vinet}}, \bibinfo {author} {\bibfnamefont
  {M.}~\bibnamefont {Urdampilleta}}, \bibinfo {author} {\bibfnamefont
  {T.}~\bibnamefont {Meunier}}, \bibinfo {author} {\bibfnamefont
  {M.}~\bibnamefont {Sanquer}}, \bibinfo {author} {\bibfnamefont
  {X.}~\bibnamefont {Jehl}}, \bibinfo {author} {\bibfnamefont {R.}~\bibnamefont
  {Maurand}}, \ and\ \bibinfo {author} {\bibfnamefont {S.}~\bibnamefont {{De
  Franceschi}}},\ }\href {\doibase 10.1038/s41467-019-10848-z} {\bibfield
  {journal} {\bibinfo  {journal} {Nat. Comm.}\ }\textbf {\bibinfo {volume}
  {10}},\ \bibinfo {pages} {2776} (\bibinfo {year} {2019})}\BibitemShut
  {NoStop}%
\bibitem [{\citenamefont {Scappucci}\ \emph {et~al.}(2020)\citenamefont
  {Scappucci}, \citenamefont {Kloeffel}, \citenamefont {Zwanenburg},
  \citenamefont {Loss}, \citenamefont {Myronov}, \citenamefont {Zhang},
  \citenamefont {Franceschi}, \citenamefont {Katsaros},\ and\ \citenamefont
  {Veldhorst}}]{Scappucci2020}%
  \BibitemOpen
  \bibfield  {author} {\bibinfo {author} {\bibfnamefont {G.}~\bibnamefont
  {Scappucci}}, \bibinfo {author} {\bibfnamefont {C.}~\bibnamefont {Kloeffel}},
  \bibinfo {author} {\bibfnamefont {F.~A.}\ \bibnamefont {Zwanenburg}},
  \bibinfo {author} {\bibfnamefont {D.}~\bibnamefont {Loss}}, \bibinfo {author}
  {\bibfnamefont {M.}~\bibnamefont {Myronov}}, \bibinfo {author} {\bibfnamefont
  {J.-J.}\ \bibnamefont {Zhang}}, \bibinfo {author} {\bibfnamefont {S.~D.}\
  \bibnamefont {Franceschi}}, \bibinfo {author} {\bibfnamefont
  {G.}~\bibnamefont {Katsaros}}, \ and\ \bibinfo {author} {\bibfnamefont
  {M.}~\bibnamefont {Veldhorst}},\ }\href {\doibase 10.1038/s41578-020-00262-z}
  {\bibfield  {journal} {\bibinfo  {journal} {Nat. Rev. Mat.}\ }\textbf
  {\bibinfo {volume} {6}},\ \bibinfo {pages} {926} (\bibinfo {year}
  {2020})}\BibitemShut {NoStop}%
\bibitem [{\citenamefont {Jirovec}\ \emph {et~al.}(2021)\citenamefont
  {Jirovec}, \citenamefont {Hofmann}, \citenamefont {Ballabio}, \citenamefont
  {Mutter}, \citenamefont {Tavani}, \citenamefont {Botifoll}, \citenamefont
  {Crippa}, \citenamefont {Kukucka}, \citenamefont {Sagi}, \citenamefont
  {Martins}, \citenamefont {Saez-Mollejo}, \citenamefont {Prieto},
  \citenamefont {Borovkov}, \citenamefont {Arbiol}, \citenamefont {Chrastina},
  \citenamefont {Isella},\ and\ \citenamefont {Katsaros}}]{Jirovec2021}%
  \BibitemOpen
  \bibfield  {author} {\bibinfo {author} {\bibfnamefont {D.}~\bibnamefont
  {Jirovec}}, \bibinfo {author} {\bibfnamefont {A.}~\bibnamefont {Hofmann}},
  \bibinfo {author} {\bibfnamefont {A.}~\bibnamefont {Ballabio}}, \bibinfo
  {author} {\bibfnamefont {P.~M.}\ \bibnamefont {Mutter}}, \bibinfo {author}
  {\bibfnamefont {G.}~\bibnamefont {Tavani}}, \bibinfo {author} {\bibfnamefont
  {M.}~\bibnamefont {Botifoll}}, \bibinfo {author} {\bibfnamefont
  {A.}~\bibnamefont {Crippa}}, \bibinfo {author} {\bibfnamefont
  {J.}~\bibnamefont {Kukucka}}, \bibinfo {author} {\bibfnamefont
  {O.}~\bibnamefont {Sagi}}, \bibinfo {author} {\bibfnamefont {F.}~\bibnamefont
  {Martins}}, \bibinfo {author} {\bibfnamefont {J.}~\bibnamefont
  {Saez-Mollejo}}, \bibinfo {author} {\bibfnamefont {I.}~\bibnamefont
  {Prieto}}, \bibinfo {author} {\bibfnamefont {M.}~\bibnamefont {Borovkov}},
  \bibinfo {author} {\bibfnamefont {J.}~\bibnamefont {Arbiol}}, \bibinfo
  {author} {\bibfnamefont {D.}~\bibnamefont {Chrastina}}, \bibinfo {author}
  {\bibfnamefont {G.}~\bibnamefont {Isella}}, \ and\ \bibinfo {author}
  {\bibfnamefont {G.}~\bibnamefont {Katsaros}},\ }\href {\doibase
  10.1038/s41563-021-01022-2} {\bibfield  {journal} {\bibinfo  {journal} {Nat.
  Mat.}\ }\textbf {\bibinfo {volume} {20}},\ \bibinfo {pages} {1106} (\bibinfo
  {year} {2021})}\BibitemShut {NoStop}%
\bibitem [{\citenamefont {Lawrie}\ \emph {et~al.}(2021)\citenamefont {Lawrie},
  \citenamefont {Russ}, \citenamefont {{van Riggelen}}, \citenamefont
  {Hendrickx}, \citenamefont {{de Snoo}}, \citenamefont {Sammak}, \citenamefont
  {Scappucci},\ and\ \citenamefont {Veldhorst}}]{Lawrie2021}%
  \BibitemOpen
  \bibfield  {author} {\bibinfo {author} {\bibfnamefont {W.~I.~L.}\
  \bibnamefont {Lawrie}}, \bibinfo {author} {\bibfnamefont {M.}~\bibnamefont
  {Russ}}, \bibinfo {author} {\bibfnamefont {F.}~\bibnamefont {{van
  Riggelen}}}, \bibinfo {author} {\bibfnamefont {N.~W.}\ \bibnamefont
  {Hendrickx}}, \bibinfo {author} {\bibfnamefont {S.~L.}\ \bibnamefont {{de
  Snoo}}}, \bibinfo {author} {\bibfnamefont {A.}~\bibnamefont {Sammak}},
  \bibinfo {author} {\bibfnamefont {G.}~\bibnamefont {Scappucci}}, \ and\
  \bibinfo {author} {\bibfnamefont {M.}~\bibnamefont {Veldhorst}},\ }\href
  {https://arxiv.org/abs/2109.07837} {\bibfield  {journal} {\bibinfo  {journal}
  {arXiv:2109.07837}\ } (\bibinfo {year} {2021})}\BibitemShut {NoStop}%
\bibitem [{\citenamefont {Hendrickx}\ \emph {et~al.}(2021)\citenamefont
  {Hendrickx}, \citenamefont {Lawrie}, \citenamefont {Russ}, \citenamefont
  {{van Riggelen}}, \citenamefont {{de Snoo}}, \citenamefont {Schouten},
  \citenamefont {Sammak}, \citenamefont {Scappucci},\ and\ \citenamefont
  {Veldhorst}}]{Hendrickx2021}%
  \BibitemOpen
  \bibfield  {author} {\bibinfo {author} {\bibfnamefont {N.~W.}\ \bibnamefont
  {Hendrickx}}, \bibinfo {author} {\bibfnamefont {W.~I.~L.}\ \bibnamefont
  {Lawrie}}, \bibinfo {author} {\bibfnamefont {M.}~\bibnamefont {Russ}},
  \bibinfo {author} {\bibfnamefont {F.}~\bibnamefont {{van Riggelen}}},
  \bibinfo {author} {\bibfnamefont {S.~L.}\ \bibnamefont {{de Snoo}}}, \bibinfo
  {author} {\bibfnamefont {R.~N.}\ \bibnamefont {Schouten}}, \bibinfo {author}
  {\bibfnamefont {A.}~\bibnamefont {Sammak}}, \bibinfo {author} {\bibfnamefont
  {G.}~\bibnamefont {Scappucci}}, \ and\ \bibinfo {author} {\bibfnamefont
  {M.}~\bibnamefont {Veldhorst}},\ }\href {\doibase 10.1038/s41586-021-03332-6}
  {\bibfield  {journal} {\bibinfo  {journal} {Nature}\ }\textbf {\bibinfo
  {volume} {591}},\ \bibinfo {pages} {580} (\bibinfo {year}
  {2021})}\BibitemShut {NoStop}%
\bibitem [{\citenamefont {Watzinger}\ \emph {et~al.}(2016)\citenamefont
  {Watzinger}, \citenamefont {Kloeffel}, \citenamefont {Vukušić},
  \citenamefont {Rossell}, \citenamefont {Sessi}, \citenamefont {Kukučka},
  \citenamefont {Kirchschlager}, \citenamefont {Lausecker}, \citenamefont
  {Truhlar}, \citenamefont {Glaser}, \citenamefont {Rastelli}, \citenamefont
  {Fuhrer}, \citenamefont {Loss},\ and\ \citenamefont
  {Katsaros}}]{watzinger2016}%
  \BibitemOpen
  \bibfield  {author} {\bibinfo {author} {\bibfnamefont {H.}~\bibnamefont
  {Watzinger}}, \bibinfo {author} {\bibfnamefont {C.}~\bibnamefont {Kloeffel}},
  \bibinfo {author} {\bibfnamefont {L.}~\bibnamefont {Vukušić}}, \bibinfo
  {author} {\bibfnamefont {M.~D.}\ \bibnamefont {Rossell}}, \bibinfo {author}
  {\bibfnamefont {V.}~\bibnamefont {Sessi}}, \bibinfo {author} {\bibfnamefont
  {J.}~\bibnamefont {Kukučka}}, \bibinfo {author} {\bibfnamefont
  {R.}~\bibnamefont {Kirchschlager}}, \bibinfo {author} {\bibfnamefont
  {E.}~\bibnamefont {Lausecker}}, \bibinfo {author} {\bibfnamefont
  {A.}~\bibnamefont {Truhlar}}, \bibinfo {author} {\bibfnamefont
  {M.}~\bibnamefont {Glaser}}, \bibinfo {author} {\bibfnamefont
  {A.}~\bibnamefont {Rastelli}}, \bibinfo {author} {\bibfnamefont
  {A.}~\bibnamefont {Fuhrer}}, \bibinfo {author} {\bibfnamefont
  {D.}~\bibnamefont {Loss}}, \ and\ \bibinfo {author} {\bibfnamefont
  {G.}~\bibnamefont {Katsaros}},\ }\href {\doibase
  10.1021/acs.nanolett.6b02715} {\bibfield  {journal} {\bibinfo  {journal}
  {Nano Lett.}\ }\textbf {\bibinfo {volume} {16}},\ \bibinfo {pages} {6879}
  (\bibinfo {year} {2016})}\BibitemShut {NoStop}%
\bibitem [{\citenamefont {Brauns}\ \emph {et~al.}(2016)\citenamefont {Brauns},
  \citenamefont {Ridderbos}, \citenamefont {Li}, \citenamefont {Bakkers},\ and\
  \citenamefont {Zwanenburg}}]{Brauns2016}%
  \BibitemOpen
  \bibfield  {author} {\bibinfo {author} {\bibfnamefont {M.}~\bibnamefont
  {Brauns}}, \bibinfo {author} {\bibfnamefont {J.}~\bibnamefont {Ridderbos}},
  \bibinfo {author} {\bibfnamefont {A.}~\bibnamefont {Li}}, \bibinfo {author}
  {\bibfnamefont {E.~P. A.~M.}\ \bibnamefont {Bakkers}}, \ and\ \bibinfo
  {author} {\bibfnamefont {F.~A.}\ \bibnamefont {Zwanenburg}},\ }\href
  {\doibase 10.1103/PhysRevB.93.121408} {\bibfield  {journal} {\bibinfo
  {journal} {Phys. Rev. B}\ }\textbf {\bibinfo {volume} {93}},\ \bibinfo
  {pages} {121408(R)} (\bibinfo {year} {2016})}\BibitemShut {NoStop}%
\bibitem [{\citenamefont {Bogan}\ \emph {et~al.}(2017)\citenamefont {Bogan},
  \citenamefont {Studenikin}, \citenamefont {Korkusinski}, \citenamefont
  {Aers}, \citenamefont {Gaudreau}, \citenamefont {Zawadzki}, \citenamefont
  {Sachrajda}, \citenamefont {Tracy}, \citenamefont {Reno},\ and\ \citenamefont
  {Hargett}}]{Bogan2017}%
  \BibitemOpen
  \bibfield  {author} {\bibinfo {author} {\bibfnamefont {A.}~\bibnamefont
  {Bogan}}, \bibinfo {author} {\bibfnamefont {S.~A.}\ \bibnamefont
  {Studenikin}}, \bibinfo {author} {\bibfnamefont {M.}~\bibnamefont
  {Korkusinski}}, \bibinfo {author} {\bibfnamefont {G.~C.}\ \bibnamefont
  {Aers}}, \bibinfo {author} {\bibfnamefont {L.}~\bibnamefont {Gaudreau}},
  \bibinfo {author} {\bibfnamefont {P.}~\bibnamefont {Zawadzki}}, \bibinfo
  {author} {\bibfnamefont {A.~S.}\ \bibnamefont {Sachrajda}}, \bibinfo {author}
  {\bibfnamefont {L.~A.}\ \bibnamefont {Tracy}}, \bibinfo {author}
  {\bibfnamefont {J.~L.}\ \bibnamefont {Reno}}, \ and\ \bibinfo {author}
  {\bibfnamefont {T.~W.}\ \bibnamefont {Hargett}},\ }\href {\doibase
  10.1103/physrevlett.118.167701} {\bibfield  {journal} {\bibinfo  {journal}
  {Phys. Rev. Lett.}\ }\textbf {\bibinfo {volume} {118}},\ \bibinfo {pages}
  {167701} (\bibinfo {year} {2017})}\BibitemShut {NoStop}%
\bibitem [{\citenamefont {Lu}\ \emph {et~al.}(2017)\citenamefont {Lu},
  \citenamefont {Harris}, \citenamefont {Huang}, \citenamefont {Chuang},
  \citenamefont {Li},\ and\ \citenamefont {Liu}}]{Lu2017}%
  \BibitemOpen
  \bibfield  {author} {\bibinfo {author} {\bibfnamefont {T.~M.}\ \bibnamefont
  {Lu}}, \bibinfo {author} {\bibfnamefont {C.~T.}\ \bibnamefont {Harris}},
  \bibinfo {author} {\bibfnamefont {S.-H.}\ \bibnamefont {Huang}}, \bibinfo
  {author} {\bibfnamefont {Y.}~\bibnamefont {Chuang}}, \bibinfo {author}
  {\bibfnamefont {J.-Y.}\ \bibnamefont {Li}}, \ and\ \bibinfo {author}
  {\bibfnamefont {C.~W.}\ \bibnamefont {Liu}},\ }\href {\doibase
  10.1063/1.4990569} {\bibfield  {journal} {\bibinfo  {journal} {Appl. Phys.
  Lett.}\ }\textbf {\bibinfo {volume} {111}},\ \bibinfo {pages} {102108}
  (\bibinfo {year} {2017})}\BibitemShut {NoStop}%
\bibitem [{\citenamefont {Crippa}\ \emph {et~al.}(2018)\citenamefont {Crippa},
  \citenamefont {Maurand}, \citenamefont {Bourdet}, \citenamefont
  {Kotekar-Patil}, \citenamefont {Amisse}, \citenamefont {Jehl}, \citenamefont
  {Sanquer}, \citenamefont {Lavi\'eville}, \citenamefont {Bohuslavskyi},
  \citenamefont {Hutin}, \citenamefont {Barraud}, \citenamefont {Vinet},
  \citenamefont {Niquet},\ and\ \citenamefont {De~Franceschi}}]{Crippa2018}%
  \BibitemOpen
  \bibfield  {author} {\bibinfo {author} {\bibfnamefont {A.}~\bibnamefont
  {Crippa}}, \bibinfo {author} {\bibfnamefont {R.}~\bibnamefont {Maurand}},
  \bibinfo {author} {\bibfnamefont {L.}~\bibnamefont {Bourdet}}, \bibinfo
  {author} {\bibfnamefont {D.}~\bibnamefont {Kotekar-Patil}}, \bibinfo {author}
  {\bibfnamefont {A.}~\bibnamefont {Amisse}}, \bibinfo {author} {\bibfnamefont
  {X.}~\bibnamefont {Jehl}}, \bibinfo {author} {\bibfnamefont {M.}~\bibnamefont
  {Sanquer}}, \bibinfo {author} {\bibfnamefont {R.}~\bibnamefont
  {Lavi\'eville}}, \bibinfo {author} {\bibfnamefont {H.}~\bibnamefont
  {Bohuslavskyi}}, \bibinfo {author} {\bibfnamefont {L.}~\bibnamefont {Hutin}},
  \bibinfo {author} {\bibfnamefont {S.}~\bibnamefont {Barraud}}, \bibinfo
  {author} {\bibfnamefont {M.}~\bibnamefont {Vinet}}, \bibinfo {author}
  {\bibfnamefont {Y.-M.}\ \bibnamefont {Niquet}}, \ and\ \bibinfo {author}
  {\bibfnamefont {S.}~\bibnamefont {De~Franceschi}},\ }\href {\doibase
  10.1103/PhysRevLett.120.137702} {\bibfield  {journal} {\bibinfo  {journal}
  {Phys. Rev. Lett.}\ }\textbf {\bibinfo {volume} {120}},\ \bibinfo {pages}
  {137702} (\bibinfo {year} {2018})}\BibitemShut {NoStop}%
\bibitem [{\citenamefont {Marcellina}\ \emph {et~al.}(2018)\citenamefont
  {Marcellina}, \citenamefont {Srinivasan}, \citenamefont {Miserev},
  \citenamefont {Croxall}, \citenamefont {Ritchie}, \citenamefont {Farrer},
  \citenamefont {Sushkov}, \citenamefont {Culcer},\ and\ \citenamefont
  {Hamilton}}]{Marcellina2018}%
  \BibitemOpen
  \bibfield  {author} {\bibinfo {author} {\bibfnamefont {E.}~\bibnamefont
  {Marcellina}}, \bibinfo {author} {\bibfnamefont {A.}~\bibnamefont
  {Srinivasan}}, \bibinfo {author} {\bibfnamefont {D.~S.}\ \bibnamefont
  {Miserev}}, \bibinfo {author} {\bibfnamefont {A.~F.}\ \bibnamefont
  {Croxall}}, \bibinfo {author} {\bibfnamefont {D.~A.}\ \bibnamefont
  {Ritchie}}, \bibinfo {author} {\bibfnamefont {I.}~\bibnamefont {Farrer}},
  \bibinfo {author} {\bibfnamefont {O.~P.}\ \bibnamefont {Sushkov}}, \bibinfo
  {author} {\bibfnamefont {D.}~\bibnamefont {Culcer}}, \ and\ \bibinfo {author}
  {\bibfnamefont {A.~R.}\ \bibnamefont {Hamilton}},\ }\href {\doibase
  10.1103/physrevlett.121.077701} {\bibfield  {journal} {\bibinfo  {journal}
  {Phys. Rev. Lett.}\ }\textbf {\bibinfo {volume} {121}},\ \bibinfo {pages}
  {077701} (\bibinfo {year} {2018})}\BibitemShut {NoStop}%
\bibitem [{\citenamefont {Gradl}\ \emph {et~al.}(2018)\citenamefont {Gradl},
  \citenamefont {Winkler}, \citenamefont {Kempf}, \citenamefont {Holler},
  \citenamefont {Schuh}, \citenamefont {Bougeard}, \citenamefont
  {Hern{\'{a}}ndez-M{\'{i}}nguez}, \citenamefont {Biermann}, \citenamefont
  {Santos}, \citenamefont {Sch{\"{u}}ller},\ and\ \citenamefont
  {Korn}}]{Gradl2018}%
  \BibitemOpen
  \bibfield  {author} {\bibinfo {author} {\bibfnamefont {C.}~\bibnamefont
  {Gradl}}, \bibinfo {author} {\bibfnamefont {R.}~\bibnamefont {Winkler}},
  \bibinfo {author} {\bibfnamefont {M.}~\bibnamefont {Kempf}}, \bibinfo
  {author} {\bibfnamefont {J.}~\bibnamefont {Holler}}, \bibinfo {author}
  {\bibfnamefont {D.}~\bibnamefont {Schuh}}, \bibinfo {author} {\bibfnamefont
  {D.}~\bibnamefont {Bougeard}}, \bibinfo {author} {\bibfnamefont
  {A.}~\bibnamefont {Hern{\'{a}}ndez-M{\'{i}}nguez}}, \bibinfo {author}
  {\bibfnamefont {K.}~\bibnamefont {Biermann}}, \bibinfo {author}
  {\bibfnamefont {P.~V.}\ \bibnamefont {Santos}}, \bibinfo {author}
  {\bibfnamefont {C.}~\bibnamefont {Sch{\"{u}}ller}}, \ and\ \bibinfo {author}
  {\bibfnamefont {T.}~\bibnamefont {Korn}},\ }\href {\doibase
  10.1103/PhysRevX.8.021068} {\bibfield  {journal} {\bibinfo  {journal} {Phys.
  Rev. X}\ }\textbf {\bibinfo {volume} {8}},\ \bibinfo {pages} {021068}
  (\bibinfo {year} {2018})}\BibitemShut {NoStop}%
\bibitem [{\citenamefont {Hofmann}\ \emph {et~al.}(2019)\citenamefont
  {Hofmann}, \citenamefont {Jirovec}, \citenamefont {Borovkov}, \citenamefont
  {Prieto}, \citenamefont {Ballabio}, \citenamefont {Frigerio}, \citenamefont
  {Chrastina}, \citenamefont {Isella},\ and\ \citenamefont
  {Katsaros}}]{hofmann2019ArXiV}%
  \BibitemOpen
  \bibfield  {author} {\bibinfo {author} {\bibfnamefont {A.}~\bibnamefont
  {Hofmann}}, \bibinfo {author} {\bibfnamefont {D.}~\bibnamefont {Jirovec}},
  \bibinfo {author} {\bibfnamefont {M.}~\bibnamefont {Borovkov}}, \bibinfo
  {author} {\bibfnamefont {I.}~\bibnamefont {Prieto}}, \bibinfo {author}
  {\bibfnamefont {A.}~\bibnamefont {Ballabio}}, \bibinfo {author}
  {\bibfnamefont {J.}~\bibnamefont {Frigerio}}, \bibinfo {author}
  {\bibfnamefont {D.}~\bibnamefont {Chrastina}}, \bibinfo {author}
  {\bibfnamefont {G.}~\bibnamefont {Isella}}, \ and\ \bibinfo {author}
  {\bibfnamefont {G.}~\bibnamefont {Katsaros}},\ }\href
  {https://arxiv.org/abs/1910.05841} {\bibfield  {journal} {\bibinfo  {journal}
  {arXiv:1910.05841}\ } (\bibinfo {year} {2019})}\BibitemShut {NoStop}%
\bibitem [{\citenamefont {Miller}\ \emph {et~al.}(2021)\citenamefont {Miller},
  \citenamefont {Brickson}, \citenamefont {Hardy}, \citenamefont {Liu},
  \citenamefont {Li}, \citenamefont {Baczewski}, \citenamefont {Lilly},
  \citenamefont {Lu},\ and\ \citenamefont {Luhman}}]{Miller2021}%
  \BibitemOpen
  \bibfield  {author} {\bibinfo {author} {\bibfnamefont {A.~J.}\ \bibnamefont
  {Miller}}, \bibinfo {author} {\bibfnamefont {M.}~\bibnamefont {Brickson}},
  \bibinfo {author} {\bibfnamefont {W.~J.}\ \bibnamefont {Hardy}}, \bibinfo
  {author} {\bibfnamefont {C.-Y.}\ \bibnamefont {Liu}}, \bibinfo {author}
  {\bibfnamefont {J.-Y.}\ \bibnamefont {Li}}, \bibinfo {author} {\bibfnamefont
  {A.}~\bibnamefont {Baczewski}}, \bibinfo {author} {\bibfnamefont {M.~P.}\
  \bibnamefont {Lilly}}, \bibinfo {author} {\bibfnamefont {T.-M.}\ \bibnamefont
  {Lu}}, \ and\ \bibinfo {author} {\bibfnamefont {D.~R.}\ \bibnamefont
  {Luhman}},\ }\href {https://arxiv.org/abs/2102.01758} {\bibfield  {journal}
  {\bibinfo  {journal} {arXiv:2102.01758}\ } (\bibinfo {year}
  {2021})}\BibitemShut {NoStop}%
\bibitem [{\citenamefont {Froning}\ \emph
  {et~al.}(2021{\natexlab{a}})\citenamefont {Froning}, \citenamefont
  {Ran\ifmmode \check{c}\else \v{c}\fi{}i\ifmmode~\acute{c}\else \'{c}\fi{}},
  \citenamefont {Het\'enyi}, \citenamefont {Bosco}, \citenamefont {Rehmann},
  \citenamefont {Li}, \citenamefont {Bakkers}, \citenamefont {Zwanenburg},
  \citenamefont {Loss}, \citenamefont {Zumb\"uhl},\ and\ \citenamefont
  {Braakman}}]{Froning2021}%
  \BibitemOpen
  \bibfield  {author} {\bibinfo {author} {\bibfnamefont {F.~N.~M.}\
  \bibnamefont {Froning}}, \bibinfo {author} {\bibfnamefont {M.~J.}\
  \bibnamefont {Ran\ifmmode \check{c}\else \v{c}\fi{}i\ifmmode~\acute{c}\else
  \'{c}\fi{}}}, \bibinfo {author} {\bibfnamefont {B.}~\bibnamefont
  {Het\'enyi}}, \bibinfo {author} {\bibfnamefont {S.}~\bibnamefont {Bosco}},
  \bibinfo {author} {\bibfnamefont {M.~K.}\ \bibnamefont {Rehmann}}, \bibinfo
  {author} {\bibfnamefont {A.}~\bibnamefont {Li}}, \bibinfo {author}
  {\bibfnamefont {E.~P. A.~M.}\ \bibnamefont {Bakkers}}, \bibinfo {author}
  {\bibfnamefont {F.~A.}\ \bibnamefont {Zwanenburg}}, \bibinfo {author}
  {\bibfnamefont {D.}~\bibnamefont {Loss}}, \bibinfo {author} {\bibfnamefont
  {D.~M.}\ \bibnamefont {Zumb\"uhl}}, \ and\ \bibinfo {author} {\bibfnamefont
  {F.~R.}\ \bibnamefont {Braakman}},\ }\href {\doibase
  10.1103/PhysRevResearch.3.013081} {\bibfield  {journal} {\bibinfo  {journal}
  {Phys. Rev. Research}\ }\textbf {\bibinfo {volume} {3}},\ \bibinfo {pages}
  {013081} (\bibinfo {year} {2021}{\natexlab{a}})}\BibitemShut {NoStop}%
\bibitem [{\citenamefont {Liles}\ \emph {et~al.}(2021)\citenamefont {Liles},
  \citenamefont {Martins}, \citenamefont {Miserev}, \citenamefont {Kiselev},
  \citenamefont {Thorvaldson}, \citenamefont {Rendell}, \citenamefont {Jin},
  \citenamefont {Hudson}, \citenamefont {Veldhorst}, \citenamefont {Itoh},
  \citenamefont {Sushkov}, \citenamefont {Ladd}, \citenamefont {Dzurak},\ and\
  \citenamefont {Hamilton}}]{Liles2021Dec}%
  \BibitemOpen
  \bibfield  {author} {\bibinfo {author} {\bibfnamefont {S.~D.}\ \bibnamefont
  {Liles}}, \bibinfo {author} {\bibfnamefont {F.}~\bibnamefont {Martins}},
  \bibinfo {author} {\bibfnamefont {D.~S.}\ \bibnamefont {Miserev}}, \bibinfo
  {author} {\bibfnamefont {A.~A.}\ \bibnamefont {Kiselev}}, \bibinfo {author}
  {\bibfnamefont {I.~D.}\ \bibnamefont {Thorvaldson}}, \bibinfo {author}
  {\bibfnamefont {M.~J.}\ \bibnamefont {Rendell}}, \bibinfo {author}
  {\bibfnamefont {I.~K.}\ \bibnamefont {Jin}}, \bibinfo {author} {\bibfnamefont
  {F.~E.}\ \bibnamefont {Hudson}}, \bibinfo {author} {\bibfnamefont
  {M.}~\bibnamefont {Veldhorst}}, \bibinfo {author} {\bibfnamefont {K.~M.}\
  \bibnamefont {Itoh}}, \bibinfo {author} {\bibfnamefont {O.~P.}\ \bibnamefont
  {Sushkov}}, \bibinfo {author} {\bibfnamefont {T.~D.}\ \bibnamefont {Ladd}},
  \bibinfo {author} {\bibfnamefont {A.~S.}\ \bibnamefont {Dzurak}}, \ and\
  \bibinfo {author} {\bibfnamefont {A.~R.}\ \bibnamefont {Hamilton}},\ }\href
  {\doibase 10.1103/PhysRevB.104.235303} {\bibfield  {journal} {\bibinfo
  {journal} {Phys. Rev. B}\ }\textbf {\bibinfo {volume} {104}},\ \bibinfo
  {pages} {235303} (\bibinfo {year} {2021})}\BibitemShut {NoStop}%
\bibitem [{\citenamefont {Qvist}\ and\ \citenamefont
  {Danon}(2022)}]{Qvist2022}%
  \BibitemOpen
  \bibfield  {author} {\bibinfo {author} {\bibfnamefont {J.~H.}\ \bibnamefont
  {Qvist}}\ and\ \bibinfo {author} {\bibfnamefont {J.}~\bibnamefont {Danon}},\
  }\href {\doibase 10.1103/PhysRevB.105.075303} {\bibfield  {journal} {\bibinfo
   {journal} {Phys. Rev. B}\ }\textbf {\bibinfo {volume} {105}},\ \bibinfo
  {pages} {075303} (\bibinfo {year} {2022})}\BibitemShut {NoStop}%
\bibitem [{\citenamefont {Ares}\ \emph {et~al.}(2013)\citenamefont {Ares},
  \citenamefont {Katsaros}, \citenamefont {Golovach}, \citenamefont {Zhang},
  \citenamefont {Prager}, \citenamefont {Glazman}, \citenamefont {Schmidt},\
  and\ \citenamefont {De~Franceschi}}]{ares2013}%
  \BibitemOpen
  \bibfield  {author} {\bibinfo {author} {\bibfnamefont {N.}~\bibnamefont
  {Ares}}, \bibinfo {author} {\bibfnamefont {G.}~\bibnamefont {Katsaros}},
  \bibinfo {author} {\bibfnamefont {V.~N.}\ \bibnamefont {Golovach}}, \bibinfo
  {author} {\bibfnamefont {J.~J.}\ \bibnamefont {Zhang}}, \bibinfo {author}
  {\bibfnamefont {A.}~\bibnamefont {Prager}}, \bibinfo {author} {\bibfnamefont
  {L.~I.}\ \bibnamefont {Glazman}}, \bibinfo {author} {\bibfnamefont {O.~G.}\
  \bibnamefont {Schmidt}}, \ and\ \bibinfo {author} {\bibfnamefont
  {S.}~\bibnamefont {De~Franceschi}},\ }\href {\doibase 10.1063/1.4858959}
  {\bibfield  {journal} {\bibinfo  {journal} {Appl. Phys. Lett.}\ }\textbf
  {\bibinfo {volume} {103}},\ \bibinfo {pages} {263113} (\bibinfo {year}
  {2013})}\BibitemShut {NoStop}%
\bibitem [{\citenamefont {Kloeffel}\ \emph {et~al.}(2011)\citenamefont
  {Kloeffel}, \citenamefont {Trif},\ and\ \citenamefont {Loss}}]{Kloeffel2011}%
  \BibitemOpen
  \bibfield  {author} {\bibinfo {author} {\bibfnamefont {C.}~\bibnamefont
  {Kloeffel}}, \bibinfo {author} {\bibfnamefont {M.}~\bibnamefont {Trif}}, \
  and\ \bibinfo {author} {\bibfnamefont {D.}~\bibnamefont {Loss}},\ }\href
  {\doibase 10.1103/PhysRevB.84.195314} {\bibfield  {journal} {\bibinfo
  {journal} {Phys. Rev. B}\ }\textbf {\bibinfo {volume} {84}},\ \bibinfo
  {pages} {195314} (\bibinfo {year} {2011})}\BibitemShut {NoStop}%
\bibitem [{\citenamefont {Kloeffel}\ \emph {et~al.}(2018)\citenamefont
  {Kloeffel}, \citenamefont {Ran\v{c}i\'{c}},\ and\ \citenamefont
  {Loss}}]{Kloeffel2018}%
  \BibitemOpen
  \bibfield  {author} {\bibinfo {author} {\bibfnamefont {C.}~\bibnamefont
  {Kloeffel}}, \bibinfo {author} {\bibfnamefont {M.~J.}\ \bibnamefont
  {Ran\v{c}i\'{c}}}, \ and\ \bibinfo {author} {\bibfnamefont {D.}~\bibnamefont
  {Loss}},\ }\href {\doibase 10.1103/physrevb.97.235422} {\bibfield  {journal}
  {\bibinfo  {journal} {Phys. Rev. B}\ }\textbf {\bibinfo {volume} {97}},\
  \bibinfo {pages} {235422} (\bibinfo {year} {2018})}\BibitemShut {NoStop}%
\bibitem [{\citenamefont {Bosco}\ \emph
  {et~al.}(2021{\natexlab{a}})\citenamefont {Bosco}, \citenamefont {Benito},
  \citenamefont {Adelsberger},\ and\ \citenamefont {Loss}}]{Bosco2021}%
  \BibitemOpen
  \bibfield  {author} {\bibinfo {author} {\bibfnamefont {S.}~\bibnamefont
  {Bosco}}, \bibinfo {author} {\bibfnamefont {M.}~\bibnamefont {Benito}},
  \bibinfo {author} {\bibfnamefont {C.}~\bibnamefont {Adelsberger}}, \ and\
  \bibinfo {author} {\bibfnamefont {D.}~\bibnamefont {Loss}},\ }\href {\doibase
  10.1103/PhysRevB.104.115425} {\bibfield  {journal} {\bibinfo  {journal}
  {Phys. Rev. B}\ }\textbf {\bibinfo {volume} {104}},\ \bibinfo {pages}
  {115425} (\bibinfo {year} {2021}{\natexlab{a}})}\BibitemShut {NoStop}%
\bibitem [{\citenamefont {Bosco}\ \emph
  {et~al.}(2021{\natexlab{b}})\citenamefont {Bosco}, \citenamefont
  {Het\'{e}nyi},\ and\ \citenamefont {Loss}}]{Bosco2021a}%
  \BibitemOpen
  \bibfield  {author} {\bibinfo {author} {\bibfnamefont {S.}~\bibnamefont
  {Bosco}}, \bibinfo {author} {\bibfnamefont {B.}~\bibnamefont {Het\'{e}nyi}},
  \ and\ \bibinfo {author} {\bibfnamefont {D.}~\bibnamefont {Loss}},\ }\href
  {\doibase 10.1103/prxquantum.2.010348} {\bibfield  {journal} {\bibinfo
  {journal} {PRX Quantum}\ }\textbf {\bibinfo {volume} {2}},\ \bibinfo {pages}
  {010348} (\bibinfo {year} {2021}{\natexlab{b}})}\BibitemShut {NoStop}%
\bibitem [{\citenamefont {Froning}\ \emph
  {et~al.}(2021{\natexlab{b}})\citenamefont {Froning}, \citenamefont
  {Camenzind}, \citenamefont {Molen}, \citenamefont {Li}, \citenamefont
  {Bakkers}, \citenamefont {Zumbühl},\ and\ \citenamefont
  {Braakman}}]{Froning2021a}%
  \BibitemOpen
  \bibfield  {author} {\bibinfo {author} {\bibfnamefont {F.~N.~M.}\
  \bibnamefont {Froning}}, \bibinfo {author} {\bibfnamefont {L.~C.}\
  \bibnamefont {Camenzind}}, \bibinfo {author} {\bibfnamefont {O.~A. H. v.~d.}\
  \bibnamefont {Molen}}, \bibinfo {author} {\bibfnamefont {A.}~\bibnamefont
  {Li}}, \bibinfo {author} {\bibfnamefont {E.~P. A.~M.}\ \bibnamefont
  {Bakkers}}, \bibinfo {author} {\bibfnamefont {D.~M.}\ \bibnamefont
  {Zumbühl}}, \ and\ \bibinfo {author} {\bibfnamefont {F.~R.}\ \bibnamefont
  {Braakman}},\ }\href {\doibase 10.1038/s41565-020-00828-6} {\bibfield
  {journal} {\bibinfo  {journal} {Nat. Nanotech.}\ }\textbf {\bibinfo {volume}
  {16}},\ \bibinfo {pages} {308} (\bibinfo {year}
  {2021}{\natexlab{b}})}\BibitemShut {NoStop}%
\bibitem [{\citenamefont {Wang}\ \emph {et~al.}(2021)\citenamefont {Wang},
  \citenamefont {Marcellina}, \citenamefont {Hamilton}, \citenamefont {Cullen},
  \citenamefont {Rogge}, \citenamefont {Salfi},\ and\ \citenamefont
  {Culcer}}]{Wang2021}%
  \BibitemOpen
  \bibfield  {author} {\bibinfo {author} {\bibfnamefont {Z.}~\bibnamefont
  {Wang}}, \bibinfo {author} {\bibfnamefont {E.}~\bibnamefont {Marcellina}},
  \bibinfo {author} {\bibfnamefont {A.~R.}\ \bibnamefont {Hamilton}}, \bibinfo
  {author} {\bibfnamefont {J.~H.}\ \bibnamefont {Cullen}}, \bibinfo {author}
  {\bibfnamefont {S.}~\bibnamefont {Rogge}}, \bibinfo {author} {\bibfnamefont
  {J.}~\bibnamefont {Salfi}}, \ and\ \bibinfo {author} {\bibfnamefont
  {D.}~\bibnamefont {Culcer}},\ }\href {\doibase 10.1038/s41534-021-00386-2}
  {\bibfield  {journal} {\bibinfo  {journal} {npj Quant. Inf.}\ }\textbf
  {\bibinfo {volume} {7}},\ \bibinfo {pages} {54} (\bibinfo {year}
  {2021})}\BibitemShut {NoStop}%
\bibitem [{\citenamefont {Winkler}(2003)}]{winkler2003}%
  \BibitemOpen
  \bibfield  {author} {\bibinfo {author} {\bibfnamefont {R.}~\bibnamefont
  {Winkler}},\ }\href@noop {} {\emph {\bibinfo {title} {{Spin--Orbit Coupling
  Effects in Two-Dimensional Electron and Hole Systems}}}}\ (\bibinfo
  {publisher} {Springer Berlin Heidelberg},\ \bibinfo {year}
  {2003})\BibitemShut {NoStop}%
\bibitem [{\citenamefont {Marcellina}\ \emph {et~al.}(2017)\citenamefont
  {Marcellina}, \citenamefont {Hamilton}, \citenamefont {Winkler},\ and\
  \citenamefont {Culcer}}]{Marcellina2017}%
  \BibitemOpen
  \bibfield  {author} {\bibinfo {author} {\bibfnamefont {E.}~\bibnamefont
  {Marcellina}}, \bibinfo {author} {\bibfnamefont {A.~R.}\ \bibnamefont
  {Hamilton}}, \bibinfo {author} {\bibfnamefont {R.}~\bibnamefont {Winkler}}, \
  and\ \bibinfo {author} {\bibfnamefont {D.}~\bibnamefont {Culcer}},\ }\href
  {\doibase 10.1103/physrevb.95.075305} {\bibfield  {journal} {\bibinfo
  {journal} {Phys. Rev. B}\ }\textbf {\bibinfo {volume} {95}},\ \bibinfo
  {pages} {075305} (\bibinfo {year} {2017})}\BibitemShut {NoStop}%
\bibitem [{\citenamefont {Philippopoulos}\ \emph {et~al.}(2020)\citenamefont
  {Philippopoulos}, \citenamefont {Chesi}, \citenamefont {Culcer},\ and\
  \citenamefont {Coish}}]{Philippopoulos2020Aug}%
  \BibitemOpen
  \bibfield  {author} {\bibinfo {author} {\bibfnamefont {P.}~\bibnamefont
  {Philippopoulos}}, \bibinfo {author} {\bibfnamefont {S.}~\bibnamefont
  {Chesi}}, \bibinfo {author} {\bibfnamefont {D.}~\bibnamefont {Culcer}}, \
  and\ \bibinfo {author} {\bibfnamefont {W.~A.}\ \bibnamefont {Coish}},\ }\href
  {\doibase 10.1103/PhysRevB.102.075310} {\bibfield  {journal} {\bibinfo
  {journal} {Phys. Rev. B}\ }\textbf {\bibinfo {volume} {102}},\ \bibinfo
  {pages} {075310} (\bibinfo {year} {2020})}\BibitemShut {NoStop}%
\bibitem [{\citenamefont {Terrazos}\ \emph {et~al.}(2021)\citenamefont
  {Terrazos}, \citenamefont {Marcellina}, \citenamefont {Wang}, \citenamefont
  {Coppersmith}, \citenamefont {Friesen}, \citenamefont {Hamilton},
  \citenamefont {Hu}, \citenamefont {Koiller}, \citenamefont {Saraiva},
  \citenamefont {Culcer},\ and\ \citenamefont {Capaz}}]{Terrazos2021}%
  \BibitemOpen
  \bibfield  {author} {\bibinfo {author} {\bibfnamefont {L.~A.}\ \bibnamefont
  {Terrazos}}, \bibinfo {author} {\bibfnamefont {E.}~\bibnamefont
  {Marcellina}}, \bibinfo {author} {\bibfnamefont {Z.}~\bibnamefont {Wang}},
  \bibinfo {author} {\bibfnamefont {S.~N.}\ \bibnamefont {Coppersmith}},
  \bibinfo {author} {\bibfnamefont {M.}~\bibnamefont {Friesen}}, \bibinfo
  {author} {\bibfnamefont {A.~R.}\ \bibnamefont {Hamilton}}, \bibinfo {author}
  {\bibfnamefont {X.}~\bibnamefont {Hu}}, \bibinfo {author} {\bibfnamefont
  {B.}~\bibnamefont {Koiller}}, \bibinfo {author} {\bibfnamefont {A.~L.}\
  \bibnamefont {Saraiva}}, \bibinfo {author} {\bibfnamefont {D.}~\bibnamefont
  {Culcer}}, \ and\ \bibinfo {author} {\bibfnamefont {R.~B.}\ \bibnamefont
  {Capaz}},\ }\href {\doibase 10.1103/physrevb.103.125201} {\bibfield
  {journal} {\bibinfo  {journal} {Phys. Rev. B}\ }\textbf {\bibinfo {volume}
  {103}},\ \bibinfo {pages} {125201} (\bibinfo {year} {2021})}\BibitemShut
  {NoStop}%
\bibitem [{\citenamefont {Han}\ \emph {et~al.}(2022)\citenamefont {Han},
  \citenamefont {Chan}, \citenamefont {de~Jong}, \citenamefont {Prosko},
  \citenamefont {Badawy}, \citenamefont {Gazibegovic}, \citenamefont {Bakkers},
  \citenamefont {Kouwenhoven}, \citenamefont {Malinowski},\ and\ \citenamefont
  {Pfaff}}]{Han2022Mar}%
  \BibitemOpen
  \bibfield  {author} {\bibinfo {author} {\bibfnamefont {L.}~\bibnamefont
  {Han}}, \bibinfo {author} {\bibfnamefont {M.}~\bibnamefont {Chan}}, \bibinfo
  {author} {\bibfnamefont {D.}~\bibnamefont {de~Jong}}, \bibinfo {author}
  {\bibfnamefont {C.}~\bibnamefont {Prosko}}, \bibinfo {author} {\bibfnamefont
  {G.}~\bibnamefont {Badawy}}, \bibinfo {author} {\bibfnamefont
  {S.}~\bibnamefont {Gazibegovic}}, \bibinfo {author} {\bibfnamefont {E.~P.
  A.~M.}\ \bibnamefont {Bakkers}}, \bibinfo {author} {\bibfnamefont {L.~P.}\
  \bibnamefont {Kouwenhoven}}, \bibinfo {author} {\bibfnamefont {F.~K.}\
  \bibnamefont {Malinowski}}, \ and\ \bibinfo {author} {\bibfnamefont
  {W.}~\bibnamefont {Pfaff}},\ }\href {https://arxiv.org/abs/2203.06047}
  {\bibfield  {journal} {\bibinfo  {journal} {arXiv:2203.06047}\ } (\bibinfo
  {year} {2022})}\BibitemShut {NoStop}%
\bibitem [{\citenamefont {Wang}\ \emph {et~al.}(2016)\citenamefont {Wang},
  \citenamefont {Klochan}, \citenamefont {Hung}, \citenamefont {Culcer},
  \citenamefont {Farrer}, \citenamefont {Ritchie},\ and\ \citenamefont
  {Hamilton}}]{Wang2016Dec}%
  \BibitemOpen
  \bibfield  {author} {\bibinfo {author} {\bibfnamefont {D.~Q.}\ \bibnamefont
  {Wang}}, \bibinfo {author} {\bibfnamefont {O.}~\bibnamefont {Klochan}},
  \bibinfo {author} {\bibfnamefont {J.-T.}\ \bibnamefont {Hung}}, \bibinfo
  {author} {\bibfnamefont {D.}~\bibnamefont {Culcer}}, \bibinfo {author}
  {\bibfnamefont {I.}~\bibnamefont {Farrer}}, \bibinfo {author} {\bibfnamefont
  {D.~A.}\ \bibnamefont {Ritchie}}, \ and\ \bibinfo {author} {\bibfnamefont
  {A.~R.}\ \bibnamefont {Hamilton}},\ }\href {\doibase
  10.1021/acs.nanolett.6b03752} {\bibfield  {journal} {\bibinfo  {journal}
  {Nano Lett.}\ }\textbf {\bibinfo {volume} {16}},\ \bibinfo {pages} {7685}
  (\bibinfo {year} {2016})}\BibitemShut {NoStop}%
\bibitem [{\citenamefont {Wang}\ \emph {et~al.}(2018)\citenamefont {Wang},
  \citenamefont {Huang}, \citenamefont {Huang}, \citenamefont {Xue},
  \citenamefont {Pan}, \citenamefont {Zhao},\ and\ \citenamefont
  {Xu}}]{Wang2018Aug}%
  \BibitemOpen
  \bibfield  {author} {\bibinfo {author} {\bibfnamefont {J.-Y.}\ \bibnamefont
  {Wang}}, \bibinfo {author} {\bibfnamefont {G.-Y.}\ \bibnamefont {Huang}},
  \bibinfo {author} {\bibfnamefont {S.}~\bibnamefont {Huang}}, \bibinfo
  {author} {\bibfnamefont {J.}~\bibnamefont {Xue}}, \bibinfo {author}
  {\bibfnamefont {D.}~\bibnamefont {Pan}}, \bibinfo {author} {\bibfnamefont
  {J.}~\bibnamefont {Zhao}}, \ and\ \bibinfo {author} {\bibfnamefont
  {H.}~\bibnamefont {Xu}},\ }\href {\doibase 10.1021/acs.nanolett.8b01153}
  {\bibfield  {journal} {\bibinfo  {journal} {Nano Lett.}\ }\textbf {\bibinfo
  {volume} {18}},\ \bibinfo {pages} {4741} (\bibinfo {year}
  {2018})}\BibitemShut {NoStop}%
\bibitem [{\citenamefont {Marx}\ \emph {et~al.}(2020)\citenamefont {Marx},
  \citenamefont {Yoneda}, \citenamefont {Rubio}, \citenamefont {Stano},
  \citenamefont {Otsuka}, \citenamefont {Takeda}, \citenamefont {Li},
  \citenamefont {Yamaoka}, \citenamefont {Nakajima}, \citenamefont {Noiri},
  \citenamefont {Loss}, \citenamefont {Kodera},\ and\ \citenamefont
  {Tarucha}}]{Marx2020Mar}%
  \BibitemOpen
  \bibfield  {author} {\bibinfo {author} {\bibfnamefont {M.}~\bibnamefont
  {Marx}}, \bibinfo {author} {\bibfnamefont {J.}~\bibnamefont {Yoneda}},
  \bibinfo {author} {\bibfnamefont {{\ifmmode\acute{A}\else\'{A}\fi}.~G.}\
  \bibnamefont {Rubio}}, \bibinfo {author} {\bibfnamefont {P.}~\bibnamefont
  {Stano}}, \bibinfo {author} {\bibfnamefont {T.}~\bibnamefont {Otsuka}},
  \bibinfo {author} {\bibfnamefont {K.}~\bibnamefont {Takeda}}, \bibinfo
  {author} {\bibfnamefont {S.}~\bibnamefont {Li}}, \bibinfo {author}
  {\bibfnamefont {Y.}~\bibnamefont {Yamaoka}}, \bibinfo {author} {\bibfnamefont
  {T.}~\bibnamefont {Nakajima}}, \bibinfo {author} {\bibfnamefont
  {A.}~\bibnamefont {Noiri}}, \bibinfo {author} {\bibfnamefont
  {D.}~\bibnamefont {Loss}}, \bibinfo {author} {\bibfnamefont {T.}~\bibnamefont
  {Kodera}}, \ and\ \bibinfo {author} {\bibfnamefont {S.}~\bibnamefont
  {Tarucha}},\ }\href {https://arxiv.org/abs/2003.07079} {\bibfield  {journal}
  {\bibinfo  {journal} {arXiv:2003.07079}\ } (\bibinfo {year}
  {2020})}\BibitemShut {NoStop}%
\bibitem [{\citenamefont {Sala}\ and\ \citenamefont {Danon}(2021)}]{Sala2021}%
  \BibitemOpen
  \bibfield  {author} {\bibinfo {author} {\bibfnamefont {A.}~\bibnamefont
  {Sala}}\ and\ \bibinfo {author} {\bibfnamefont {J.}~\bibnamefont {Danon}},\
  }\href {\doibase 10.1103/physrevb.104.085421} {\bibfield  {journal} {\bibinfo
   {journal} {Phys. Rev. B}\ }\textbf {\bibinfo {volume} {104}},\ \bibinfo
  {pages} {085421} (\bibinfo {year} {2021})}\BibitemShut {NoStop}%
\bibitem [{\citenamefont {Danon}\ and\ \citenamefont
  {Nazarov}(2009)}]{danon2009}%
  \BibitemOpen
  \bibfield  {author} {\bibinfo {author} {\bibfnamefont {J.}~\bibnamefont
  {Danon}}\ and\ \bibinfo {author} {\bibfnamefont {Y.~V.}\ \bibnamefont
  {Nazarov}},\ }\href {\doibase 10.1103/physrevb.80.041301} {\bibfield
  {journal} {\bibinfo  {journal} {Phys. Rev. B}\ }\textbf {\bibinfo {volume}
  {80}},\ \bibinfo {pages} {041301} (\bibinfo {year} {2009})}\BibitemShut
  {NoStop}%
\bibitem [{\citenamefont {Jouravlev}\ and\ \citenamefont
  {Nazarov}(2006)}]{jouravlev2006}%
  \BibitemOpen
  \bibfield  {author} {\bibinfo {author} {\bibfnamefont {O.~N.}\ \bibnamefont
  {Jouravlev}}\ and\ \bibinfo {author} {\bibfnamefont {Y.~V.}\ \bibnamefont
  {Nazarov}},\ }\href {\doibase 10.1103/physrevlett.96.176804} {\bibfield
  {journal} {\bibinfo  {journal} {Phys. Rev. Lett.}\ }\textbf {\bibinfo
  {volume} {96}},\ \bibinfo {pages} {176804} (\bibinfo {year}
  {2006})}\BibitemShut {NoStop}%
\bibitem [{\citenamefont {Danon}\ \emph {et~al.}(2013)\citenamefont {Danon},
  \citenamefont {Wang},\ and\ \citenamefont {Manchon}}]{danon2013}%
  \BibitemOpen
  \bibfield  {author} {\bibinfo {author} {\bibfnamefont {J.}~\bibnamefont
  {Danon}}, \bibinfo {author} {\bibfnamefont {X.}~\bibnamefont {Wang}}, \ and\
  \bibinfo {author} {\bibfnamefont {A.}~\bibnamefont {Manchon}},\ }\href
  {\doibase 10.1103/physrevlett.111.066802} {\bibfield  {journal} {\bibinfo
  {journal} {Phys. Rev. Lett.}\ }\textbf {\bibinfo {volume} {111}},\ \bibinfo
  {pages} {066802} (\bibinfo {year} {2013})}\BibitemShut {NoStop}%
\bibitem [{Note1()}]{Note1}%
  \BibitemOpen
  \bibinfo {note} {The current also vanishes for $\theta =\pi /2$ and $\eta
  =\pi /2$, where the couplings between $|{S_{02}}\rangle $ and $\{|{\uparrow
  \downarrow }\rangle ,|{\downarrow \uparrow }\rangle \}$ (in the eigenbasis of
  the local fields) are zero. However, the case $\eta =\pi /2$ corresponds to
  the extreme case where $t_{\protect \rm so}=t$ and $t_s=0$, which we will
  ignore here.}\BibitemShut {Stop}%
\bibitem [{\citenamefont {Fischer}\ \emph {et~al.}(2008)\citenamefont
  {Fischer}, \citenamefont {Coish}, \citenamefont {Bulaev},\ and\ \citenamefont
  {Loss}}]{Fischer2008}%
  \BibitemOpen
  \bibfield  {author} {\bibinfo {author} {\bibfnamefont {J.}~\bibnamefont
  {Fischer}}, \bibinfo {author} {\bibfnamefont {W.~A.}\ \bibnamefont {Coish}},
  \bibinfo {author} {\bibfnamefont {D.~V.}\ \bibnamefont {Bulaev}}, \ and\
  \bibinfo {author} {\bibfnamefont {D.}~\bibnamefont {Loss}},\ }\href {\doibase
  10.1103/PhysRevB.78.155329} {\bibfield  {journal} {\bibinfo  {journal} {Phys.
  Rev. B}\ }\textbf {\bibinfo {volume} {78}},\ \bibinfo {pages} {155329}
  (\bibinfo {year} {2008})}\BibitemShut {NoStop}%
\bibitem [{\citenamefont {Testelin}\ \emph {et~al.}(2009)\citenamefont
  {Testelin}, \citenamefont {Bernardot}, \citenamefont {Eble},\ and\
  \citenamefont {Chamarro}}]{Testelin2009}%
  \BibitemOpen
  \bibfield  {author} {\bibinfo {author} {\bibfnamefont {C.}~\bibnamefont
  {Testelin}}, \bibinfo {author} {\bibfnamefont {F.}~\bibnamefont {Bernardot}},
  \bibinfo {author} {\bibfnamefont {B.}~\bibnamefont {Eble}}, \ and\ \bibinfo
  {author} {\bibfnamefont {M.}~\bibnamefont {Chamarro}},\ }\href {\doibase
  10.1103/physrevb.79.195440} {\bibfield  {journal} {\bibinfo  {journal} {Phys.
  Rev. B}\ }\textbf {\bibinfo {volume} {79}},\ \bibinfo {pages} {195440}
  (\bibinfo {year} {2009})}\BibitemShut {NoStop}%
\bibitem [{\citenamefont {Prechtel}\ \emph {et~al.}(2016)\citenamefont
  {Prechtel}, \citenamefont {Kuhlmann}, \citenamefont {Houel}, \citenamefont
  {Ludwig}, \citenamefont {Valentin}, \citenamefont {Wieck},\ and\
  \citenamefont {Warburton}}]{Prechtel2016}%
  \BibitemOpen
  \bibfield  {author} {\bibinfo {author} {\bibfnamefont {J.~H.}\ \bibnamefont
  {Prechtel}}, \bibinfo {author} {\bibfnamefont {A.~V.}\ \bibnamefont
  {Kuhlmann}}, \bibinfo {author} {\bibfnamefont {J.}~\bibnamefont {Houel}},
  \bibinfo {author} {\bibfnamefont {A.}~\bibnamefont {Ludwig}}, \bibinfo
  {author} {\bibfnamefont {S.~R.}\ \bibnamefont {Valentin}}, \bibinfo {author}
  {\bibfnamefont {A.~D.}\ \bibnamefont {Wieck}}, \ and\ \bibinfo {author}
  {\bibfnamefont {R.~J.}\ \bibnamefont {Warburton}},\ }\href {\doibase
  10.1038/nmat4704} {\bibfield  {journal} {\bibinfo  {journal} {Nat. Mat.}\
  }\textbf {\bibinfo {volume} {15}},\ \bibinfo {pages} {981} (\bibinfo {year}
  {2016})}\BibitemShut {NoStop}%
\bibitem [{\citenamefont {Bosco}\ and\ \citenamefont
  {Loss}(2021)}]{Bosco2021b}%
  \BibitemOpen
  \bibfield  {author} {\bibinfo {author} {\bibfnamefont {S.}~\bibnamefont
  {Bosco}}\ and\ \bibinfo {author} {\bibfnamefont {D.}~\bibnamefont {Loss}},\
  }\href {\doibase 10.1103/physrevlett.127.190501} {\bibfield  {journal}
  {\bibinfo  {journal} {Phys. Rev. Lett.}\ }\textbf {\bibinfo {volume} {127}},\
  \bibinfo {pages} {190501} (\bibinfo {year} {2021})}\BibitemShut {NoStop}%
\bibitem [{\citenamefont {Chekhovich}\ \emph {et~al.}(2013)\citenamefont
  {Chekhovich}, \citenamefont {Glazov}, \citenamefont {Krysa}, \citenamefont
  {Hopkinson}, \citenamefont {Senellart}, \citenamefont {Lemaître},
  \citenamefont {Skolnick},\ and\ \citenamefont
  {Tartakovskii}}]{Chekhovich2013}%
  \BibitemOpen
  \bibfield  {author} {\bibinfo {author} {\bibfnamefont {E.~A.}\ \bibnamefont
  {Chekhovich}}, \bibinfo {author} {\bibfnamefont {M.~M.}\ \bibnamefont
  {Glazov}}, \bibinfo {author} {\bibfnamefont {A.~B.}\ \bibnamefont {Krysa}},
  \bibinfo {author} {\bibfnamefont {M.}~\bibnamefont {Hopkinson}}, \bibinfo
  {author} {\bibfnamefont {P.}~\bibnamefont {Senellart}}, \bibinfo {author}
  {\bibfnamefont {A.}~\bibnamefont {Lemaître}}, \bibinfo {author}
  {\bibfnamefont {M.~S.}\ \bibnamefont {Skolnick}}, \ and\ \bibinfo {author}
  {\bibfnamefont {A.~I.}\ \bibnamefont {Tartakovskii}},\ }\href {\doibase
  10.1038/nphys2514} {\bibfield  {journal} {\bibinfo  {journal} {Nat. Phys.}\
  }\textbf {\bibinfo {volume} {9}},\ \bibinfo {pages} {74} (\bibinfo {year}
  {2013})}\BibitemShut {NoStop}%
\bibitem [{\citenamefont {Machnikowski}\ \emph {et~al.}(2019)\citenamefont
  {Machnikowski}, \citenamefont {Gawarecki},\ and\ \citenamefont
  {Cywi\ifmmode~\acute{n}\else \'{n}\fi{}ski}}]{Machnikowski2019}%
  \BibitemOpen
  \bibfield  {author} {\bibinfo {author} {\bibfnamefont {P.}~\bibnamefont
  {Machnikowski}}, \bibinfo {author} {\bibfnamefont {K.}~\bibnamefont
  {Gawarecki}}, \ and\ \bibinfo {author} {\bibfnamefont {L.}~\bibnamefont
  {Cywi\ifmmode~\acute{n}\else \'{n}\fi{}ski}},\ }\href {\doibase
  10.1103/physrevb.100.085305} {\bibfield  {journal} {\bibinfo  {journal}
  {Phys. Rev. B}\ }\textbf {\bibinfo {volume} {100}},\ \bibinfo {pages}
  {085305} (\bibinfo {year} {2019})}\BibitemShut {NoStop}%
\end{thebibliography}
\end{document}